\documentclass[nologo,11pt,a4paper]{ETHpaper}

\usepackage{graphicx, epsfig, amsmath, amssymb}
\usepackage[square,numbers,sort&compress]{natbib}

\begin{document}

\newcommand{\mean}[1]{\left\langle #1 \right\rangle}
\newcommand{\abs}[1]{\left| #1 \right|}

\renewcommand{\thefootnote}{ \fnsymbol{footnote} }

\title{\mbox{Risk-Seeking} versus \mbox{Risk-Avoiding Investments} in
  \mbox{Noisy Periodic Environments}}

\titlealternative{Risk-Seeking vs. Risk-Avoiding Investments in
  Noisy Periodic Environments}

\author{\mbox{J. Emeterio Navarro B.$^\star$, Frank E. Walter$^\dagger$}, Frank Schweitzer$^\dagger$\footnote{Corresponding author: \url{fschweitzer@ethz.ch}}}

\authoralternative{J. E. Navarro B., F. E. Walter, F. Schweitzer}

 \address{$^\star$Institute for Informatics, Humboldt University, 10099 Berlin, Germany \\
         $^\dagger$Chair of Systems Design, ETH  Zurich, Kreuzplatz 5, 8032 Zurich, Switzerland}

 \reference{\emph{International Journal of Modern Physics C} vol. 19, no
   6 (2008), pp. 971-994 \\ See \url{http://www.sg.ethz.ch} for more information. }

 \www{\url{http://www.sg.ethz.ch}}

\makeframing
\maketitle

\begin{abstract}
  We study the performance of various agent strategies in an artificial
  investment scenario.  Agents are equipped with a budget, $x(t)$, and at
  each time step invest a particular fraction, $q(t)$, of their budget.
  The return on investment (RoI), $r(t)$, is characterized by a periodic
  function with different types and levels of noise.  Risk-avoiding
  agents choose their fraction $q(t)$ proportional to the expected
  positive RoI, while risk-seeking agents always choose a maximum value
  $q_{max}$ if they predict the RoI to be positive (``everything on
  red''). In addition to these different strategies, agents have
  different capabilities to predict the future $r(t)$, dependent on their
  internal complexity.  Here, we compare 'zero-intelligent' agents using
  technical analysis (such as moving least squares) with agents using
  reinforcement learning or genetic algorithms to predict $r(t)$.  The
  performance of agents is measured by their average budget growth after
  a certain number of time steps. We present results of extensive
  computer simulations, which show that, for our given artificial
  environment, (i) the risk-seeking strategy outperforms the
  risk-avoiding one, and (ii) the genetic algorithm was able to find this
  optimal strategy itself, and thus outperforms other prediction
  approaches considered.
  
\emph{keywords:} {risk, investment strategies, genetic algorithm}

\end{abstract}

\section{Introduction}
\label{sec:Introduction}

In the course of this paper, we investigate a model in which agents with
different strategies participate in an simple investment scenario with
noisy returns \cite{NavarroEtAl07}.  We use this setup to approach the
question how the (internal) complexity of agents enhances their
performance in a hard-to-predict environment. In the field of artificial
intelligence and complex systems, one can distinguish between two types
of agents: first, agents which only react on external changes (also known
as ``zero-intelligence agents'' \citep{Gode-Sunder93, Farmer-Patelli04})
and second, agents which have a complex internal architecture (e.g.
``belief-desire-intention agents'').  Despite these clear differences in
agent architecture, it is difficult to determine what influence these
properties have on the overall performance of the agents. In order to
study this question in a controlled environment, we have chosen an
investment model with noisy returns, to compare the performance of simple
and complex agents.  To which extent is worthwhile to equip an agent with
complex learning mechanisms instead of having a reactive response to
exogenous returns?

As a necessary step to study more complex scenarios, we are interested in
an initial study of the performance of different agent
architectures/investment strategies in the following simple setup: each
agent has a certain budget $x(t)$ and is able to invest a certain
fraction of its budget on a market. The gain or loss it makes depends on
the market return, or return on investment (RoI). In other words, at each
time step $t$, the agent adjust its risk propensity, the fraction of its
budget that it are willing to invest on the market, denoted by $q(t)$,
thereby controlling gains and losses resulting from the RoI, denoted by
$r(t)$.  We assume that only the past and current values of $r(t)$ are
known to the agent; it does not know the dynamics governing future values
of $r(t)$.  Agents observe the market through the value of $r(t)$ and,
based on analysing a set of past $r(t)$ values, they predict future
$r(t)$ values and determine their behaviour on that market through
specifying $q(t)$.

In this simple model, we consider agents that invest independently in
the market, i.e. there is no interaction or communication with other
agents. Also, there is no feedback of the investments done by agents on
the market return.  In other words, the \emph{environment} of the agents
is not influenced by their investments. This is a crucial assumption
which makes our model different from other attempts to model real market
dynamics, e.g. as for financial markets \citep{LeBaron01, Lux-Marchesi02,
  Raberto-Cincotti03}.  Consequently, we do not construct and investigate
a market model; rather, our focus lies on investigating what are good and
what are bad strategies -- in a rather artificial and controlled market
environment (see also Section \ref{sec:ReturnOnInvestment}). Regarding
the relevance of our results for real financial markets, see also our
comments in the concluding Section \ref{sec:Conclusions}. 

\begin{figure}[htbm]
  \begin{center}
    \begin{minipage}[c]{0.49\linewidth}
\vspace{0.3cm}
      \includegraphics[width=6.7cm]{figures/roi_aaa_percent}
\vspace{0.3cm}
    \end{minipage}
    \hfill
    \begin{minipage}[c]{0.49\linewidth}
      \includegraphics[width=6.7cm]{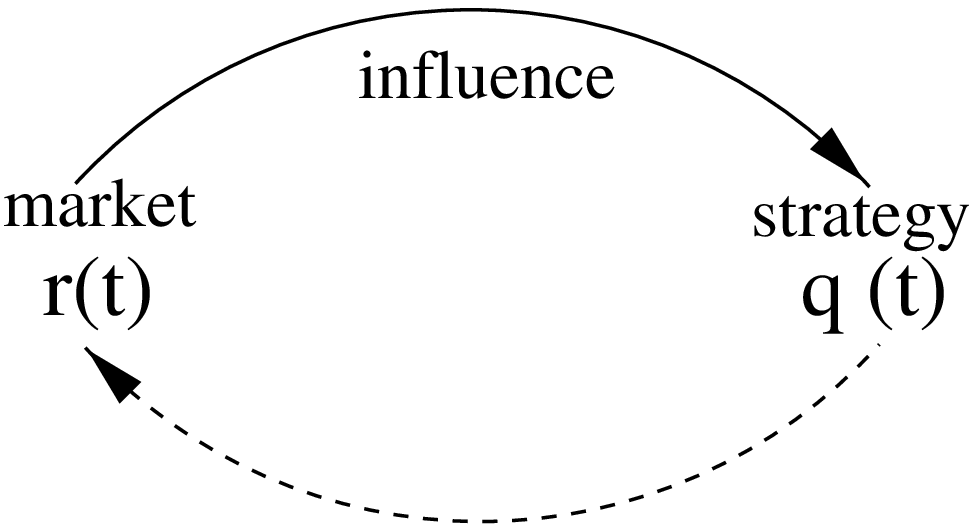}
    \end{minipage}
    \begin{minipage}[t]{1.0\linewidth}
      \begin{center}
        (a) \hspace{6cm} (b)     
      \end{center}
    \end{minipage}
    \caption{(a) Real returns of a stock item (in this example, the stock
      was AAA, the Altana Group) in percent over time (01/05/2002 --
      01/03/2004). $r(t)$ was computed for a current price $p(t)$ as
      follows: $r(t)=\log{p(t)}-\log{p(t-1)}$. (b) The dynamics of the
      market as indicated by the market return $r(t)$ and the strategy as
      defined by the invested budget $q(t)$. Note that our model does not
      consider the feedback of investments on the market dynamics (dashed
      line).}
    \label{fig:rq}
  \end{center}
\end{figure}

The essence of the model is captured in Figure \ref{fig:rq}: (a) plots
the returns in percent of a real stock item over the range of about two
years.  This illustrates the range and shape of values of returns of a
real-world stock item. (b) illustrates the dynamics of the model: $r(t)$,
the market return, influences the strategies agents have to adjust
$q(t)$, the risk propensity.  In this particular model, we do not
consider the influence that adjusted risk propensity has on the market
return, i.e. the influence of $q(t)$ on $r(t)$.

The challenge for the agents thus is twofold: \emph{first}, agents have
to predict $r(t)$ as accurately as possible, and \emph{second}, they have
to adjust $q(t)$ to the proper values as quickly as possible. This is a
complex and difficult task since most investment environments are
uncertain and fluctuating. Choosing to avoid risk and investing too
little may lead to small gains, and choosing to take risk and investing
too much may lead to large losses.

The task of finding an appropriate strategy that controls the risk and
balances between these two extrema is by far not trivial.  Methods from
technical analysis, such as estimations based on \emph{moving averages}
or \emph{moving least squares} (see also Section
\ref{sec:TechnicalAnalysis}) try to approximate the behaviour of the
environment and, based on that approximation, determine the most
appropriate investment at a particular time.  In the theory of risk,
several authors assume that individuals choose among assets based on the
mean return and on the variance of the return \citep{markowitz91,
  maslov98, kirman05}. Others have focused their attention to the
important task of how to measure risk, which lead to different type of
measures \citep{pratt64, Artzner99}.  In general, these measures are
based on the risk aversion of a decision maker having the choice to
receive a random or a non-random amount.

A typical scenario to study investment strategies is to let an agent
choose between investing in a risk-free asset or in a risky
asset\citep{tobin58,pratt64}.  It was shown\citep{tobin58} that sometimes
it may be more reasonable to invest in a risk-free asset as a means to
transfer wealth over time.  However, assuming a model with no consumption
\citep{hens06}, those agents investing in risk-free assets will be driven
out of the market in the long run by agents investing in a risky asset.
When dealing with risky assets, it is typically assumed that the agent
considers the expected return and its volatility as indicators for the
investment strategy \citep{markowitz91}.

For the sake of simplicity, in this paper we assume that the agent's
behavior is risk-neutral in the sense that the agent estimates only the
expected return, $r(t)$, and does not consider risk measures such as the
volatility.  Based on the estimation of $r(t)$, the decision to increase
or decrease the investment fraction of the risky asset should be taken.
Hence, the two terms 'risk-seeking' and 'risk-avoiding' refer only to the
choice of the investment fraction, $q(t)$.

The remainder of this paper is organised as follows: In the next section,
Section \ref{sec:Model}, we present the details of the investment model,
the properties and abilities of an agent acting/investing in this
environment.  Following that, Section \ref{sec:AgentStrategies} presents
some of the strategies that an agent can use to control its risk
propensity and Section \ref{sec:ReturnOnInvestment} presents the
properties of the return on investment (RoI) that we are considering;
Section \ref{sec:OptimalParameterAdjustment} illustrates the optimal
parameter adjustment for the presented strategies and their derivation;
Section \ref{sec:Results} compares the different investment strategies.
This is done by means of simulations where the average total budget of
agents using each investment strategy is obtained for a large a number of
trials. Finally, in Section \ref{sec:Conclusions}, we present our
conclusions.

\section{Model}
\label{sec:Model}

In this model, agents are characterised by the following two variables:
\begin{enumerate}
\item their \emph{budget} $x(t)$, which is a measure of their ``wealth''
  or ``liquidity'', and
\item the \emph{strategy} that they employ in order to control the
  fraction of the budget $q(t)$ to invest at each time step.
\end{enumerate}
In other words, at each time step $t$, an agent invests a portion
$q(t)x(t)$ of its budget. The investment yields a gain or a loss,
determined by the value of $r(t)$. Being a fraction of the investment
budget, $q(t)$, is, of course, restricted to the interval between
$[0,1]$. However, in our model we further restrict it to be from the
interval $[q_{min},q_{max}]$ where we choose $q_{min}=0.1$ and
$q_{max}=1.0$.  This implies that, at each time step $t$, there is a
minimal investment of $0.1$ of the budget, and a maximal investment of
the entire budget.

We can then define the dynamics for the budget of agents $x(t)$ as
\begin{equation}
  \label{eq:budget}
  x(t+1)=x(t)\,\Big[1+r(t)\,q(t)\Big]
\end{equation}
where $r(t)$ is the market return at the previous time step $t$.  The
market return function $r(t)$ is restricted to the range of $[-1,1]$.  A
value of $r(t)=-1$ corresponds to a total loss of the invested fraction
of the budget $q(t)$ and $r(t)=1$ corresponds to a gain equivalent to the
invested fraction.  Thus, an agent can, at any time step $t$, loose its
complete budget (for $q(t)=1$ and $r(t)=-1$), but also double its budget
(for $q(t)=1$ and $r(t)=1$). In principle, there is no upper boundary for
$r(t)$, $r(t)=1$ was chosen to obtain a mean of zero for $r(t)$ which
allows us to better understand the basic dynamics of this model. We
emphasize again that, in our model, the aim is not a most realistic
simulation of the market return, but a comparison of different agent
strategies.  The difficulty for the agents lies in properly predicting
the next value of $r(t)$ and then adjusting $q(t)$ fast enough.

Note that the restriction of $r(t)$ to the range of $[-1,1]$ is not a
realistic assumption for a real market. There, some $r(t)$ will also fall
into the range $)-1,1($ -- these are rare, extreme events that occur,
e.g. in cases of stock market bubbles and crashes. However, normally,
returns will be in the range of $[-1,1]$, e.g. as the ones of the stock
depicted in Figure \ref{fig:rq}. As we would like to focus on the
questions of choosing appropriate agent strategies in environments with
noisy, periodic returns, it is reasonable to exclude such rare, extreme
events and assume a restriction of $r(t)$ to the range of $[-1,1]$.

In the next two sections, we will outline the agent strategies (Section
\ref{sec:AgentStrategies}) and return on investment (Section
\ref{sec:ReturnOnInvestment}) that we consider.

\section{Agent Strategies}
\label{sec:AgentStrategies}

As explained before, we are interested in how the market dynamics,
$r(t)$, affect the different investment strategies of the agent, $q(t)$.
It is very important to realise that the market dynamics -- while
affecting each agent's $q(t)$ -- are \emph{not known} to the agents.
I.e., at time $t$, each agent only receives the \emph{actual value} of
the Return on Investment (RoI) and adjusts its risk propensity
accordingly, \emph{without} having a complete knowledge about the
dynamics of $r(t)$. The agent may, of course, have some bounded memory
about past RoI that could be used for predictions of future RoI. However,
the agent has to gather information about the ups and downs of the RoI
and to draw its own conclusions from this information by itself.
Therefore, the agent will perform better in the environment if it is able
to guess the market dynamics.

In the following, we present a selection of strategies that can be
applied by agents. We distinguish a reference strategy, which serves as a
frame of reference to compare and evaluate the performance of other
strategies, as well as technical analysis-based and machine
learning-based strategies.  Usually (there are exceptions, as will be
discussed in the following), a strategy consists of two components: a
\emph{prediction component} and an \emph{action component}.  For such
strategies, the prediction component predicts a variable in the system --
in this case, the next value of $r(t)$ -- and the action component then
defines an action upon the prediction of the variable -- in this case, it
defines the appropriate value for $q(t)$.

\subsection{Reference Strategy}
\label{sec:ReferenceStrategy}

In order to compare different strategies, we need a point of reference
against which the performance of each strategy can be measured. The
reference strategy that we are using is the most simple strategy
possible, i.e. that an agent always assumes a \emph{constant
  risk-propensity value} $q_0$ at every time step $t$:
\begin{equation}
  \label{eq:q0} q(t)=q_0=\mathrm{const.}
\end{equation}
Since the value of $q(t)$ is always fixed, this is not really a
``strategy'', but it plays a role in more physics-inspired investment
models \citep{Sornette-Cont97,Takayasu-Sato97,NavarroEtAl07}. We use this
strategy to compare it with more complex strategies. Note that this
reference strategy requires no knowledge on the RoI.

\subsection{Technical Analysis}
\label{sec:TechnicalAnalysis}

The following simple strategies for risk adjustment are based on
``technical analysis'' \citep{Brooks02}. Technical analysis tries to
deduce information about the dynamics of $r(t)$ by looking at trends
(averages, variances, higher order moments) of the RoI values over a
range of time. This assumes that an agent has a bounded memory of size
$M$ to record previous RoI; this information is then processed in
different ways to predict the next RoI. In the following, we consider two
strategies from the field of technical analysis: the first strategy is
based on calculating \emph{moving averages} (MA) on previous RoI, while
the second strategy uses \emph{moving least squares} (MLS) on previous
RoI, $r(t)$, over a fixed period of time, $M$. Both of them can be
regarded as ``zero-intelligence'' strategies, as agents do not do any
reasoning or learning. 

\subsubsection{Moving Averages}

The moving averages technique computes $\hat{r}_{MA}(t)$, an estimate of
the next $r(t)$, as the average of the previous $M$ values of $r(t)$:
\begin{equation}
  \label{eq:ma} \hat{r}_{MA}(t)=\frac{1}{M}\sum_{n=t-M}^{t-1}r(n)
\end{equation}

\subsubsection{Moving Least Squares}

The moving least squares technique fits a function to the data of the
previous $M$ values of $r(t)$ to estimate the next $r(t)$. In our case,
we choose this function to be a linear trend-line, which is found by
minimising the distance to the data points of $r(t)$. Based on the
previous $M$ values of $r(t)$, the squared estimation error
$\epsilon_{r}$ is defined as:
\begin{equation}
  \label{eq:mls_error} 
\epsilon_{r}(t)=\frac{1}{M} \sum\limits_{n=t-M+1}^{t}\left[r(n) -
  \hat{r}_{MLS}(n)\right]^{2} 
\end{equation}
where $\hat{r}_{MLS}(t)$ is the predicted RoI based on the linear
regression trend-line, defined as:
\begin{equation}
  \label{eq:mls_regression} 
  \hat{r}_{MLS}(t^{\prime})=m(t)\,t^{\prime}+b(t) \quad \mathrm{for}\;
  t-M \leq t^{\prime} \leq t 
\end{equation}
Now, the best fitting values $m$ and $b$ are obtained by minimising the
squared error estimation, eq. (\ref{eq:mls_error}). From $\partial
\epsilon_{r}/\partial m=0$ and $\partial \epsilon_{r}/\partial b=0$, we
get, as it is well known:
\begin{eqnarray}
  \label{eq:mls}
  m(t)&=&\frac{M\sum\limits_{n=t-M+1}^{t}n\,r(n)\;-\;\left(\sum\limits_{n=t-M+1}^{t}n\right)\left(\sum\limits_{n=t-M+1}^{t}r(n)\right)}{M\sum\limits_{n=t-M+1}^{t}n^{2}- \left(\sum\limits_{n=t-M+1}^{t}n\right)^{2}}
  \\
  b(t)&=&\frac{1}{M}\left[\sum\limits_{n=t-M+1}^{t}r(n)-m(t)\sum\limits_{n=t-M+1}^{t}n \right]
\end{eqnarray}
These two strategies use different approaches to estimate future $r(t)$;
it remains to define the corresponding adjustment of the risk propensity:
here, we consider two possibilities. First, a \emph{risk-seeking} (RS)
and, second, a \emph{risk-avoiding} (RA) approach.  In the risk-seeking
approach, the value of $q_{RS}(t)$ is defined as follows for
$\hat{r}(t)\in\{\hat{r}_{MA}(t),\hat{r}_{MLS}(t)\}$, i.e.  for
$\hat{r}(t)$ being an MA or MLS estimate of $r(t)$:
\begin{equation}
  \label{eq:mamls_rs}
  q_{RS}(t) =
  \begin{cases}
    q_{\mathrm{\min}} & \textrm{$\hat{r}(t) \le 0$} \\ q_{\mathrm{\max}}
    & \textrm{$\hat{r}(t) > 0$}
  \end{cases}
\end{equation}
where $q_{\mathrm{\min}}, q_{\mathrm{\max}} \in [0,1]$ and
$q_{\mathrm{\min}} < q_{\mathrm{\max}}$. In other words, agents invest
$q_{\mathrm{\min}}$ if the next value of $r(t)$ is predicted to be
negative or zero, and agents invest $q_{\mathrm{\max}}$ if the next value
of $r(t)$ is predicted to be positive.

In the risk-avoiding approach, the value of $q_{RA}(t)$ is defined as
follows for $\hat{r}(t)\in\{\hat{r}_{MA}(t),\hat{r}_{MLS}(t)\}$, i.e.
for $\hat{r}(t)$ being an MA or MLS estimate of $r(t)$:
\begin{equation}
  \label{eq:mamls_ra}
  q_{RA}(t) =
  \begin{cases}
    q_{\mathrm{\min}} & \textrm{$\hat{r}(t) \le q_{\mathrm{\min}}$} \\
    \hat{r}(t) & q_{\mathrm{\min}} < \hat{r}(t) < q_{\mathrm{\max}}\\
    q_{\mathrm{\max}} & \textrm{$\hat{r}(t) \ge q_{\mathrm{\max}}$} 
  \end{cases}
\end{equation}
where $q_{\mathrm{\min}}, q_{\mathrm{\max}} \in [0,1]$ and
$q_{\mathrm{\min}} < q_{\mathrm{\max}}$. Here, the respective $q(t)$ is
set to the predicted $r(t)$ (with appropriate adjustments to ensure that
$q(t)=q_{min}$ whenever $\hat{r}(t) \le q_{min}$ and $q(t)=q_{max}$
whenever $\hat{r}(t) \ge q_{max}$) -- agents only invest a fraction of
the budget which corresponds in size to the expected return.

\subsection{Machine Learning Approaches}
\label{sec:MachineLearningApproaches}

Another class of strategies for risk adjustment is based on more complex
agent information processing capabilities from the field of machine
learning. In this paper, we consider two such approaches: one based on an
incremental update rule (IUR), which is a form of reinforcement learning,
and the other based on a genetic algorithm (GA), which is a form of
evolutionary learning.

\subsubsection{Incremental Update Rule}

The following machine learning approach is based on the incremental
update rule, an application of reinforcement learning \citep{Sutton98}.
The idea of reinforcement learning is that an agent continuously uses a
reward signal to adjust its own performance. In our scenario, the values
of the return are the reward signal; at each step, the agent computes the
error between the predicted and the actual value of the return and uses
this error to adjust the estimation of the following return. The general
incremental update rule from reinforcement learning is defined as
follows:
\begin{equation} 
  \label{eq:iur} \textrm{\emph{NewEst}} \gets \textrm{\emph{OldEst}} + \textrm{\emph{StepSize}} [ \textrm{\emph{Target}} - \textrm{\emph{OldEst }} ]
\end{equation}
$OldEst$ and $NewEst$ are the old and new estimates for the quantity of
interest. So, $Target-OldEst$ gives the error of the current estimation,
which is weighted by the factor $StepSize$. This is, a new estimate is
computed by taking the old estimate and adjusting it by the error of the
current estimate. $NewEst$ has to be updated at each time step.  Applying
eq.  (\ref{eq:iur}) to our model, we find the following instance of the
incremental update rule:
\begin{equation}
  \label{eq:iur_instance} \hat{r}_{IUR}(t+1) = \hat{r}_{IUR}(t) + \gamma \bigl[r(t) - \hat{r}_{IUR}(t) \bigr]
\end{equation}
Consequently, $OldEst$ and $NewEst$ are the old and new estimates for the
return, $\hat{r}_{IUR}(t)$ and $\hat{r}_{IUR}(t+1)$; furthermore, $r(t) -
\hat{r}_{IUR}(t)$ is the error of the current estimate.  Because of its
recursive definition, the incremental update rule considers an infinite
history of returns -- of course, the weight of a value depends on its age
and its impact fades over time. We chose $\hat{r}_{IUR}(0)=0$ as the
initial value of $\hat{r}_{IUR}(t)$.  Different values of $\gamma$ lead
to different performance of the algorithm; in other words, for small
$\gamma$, the adjustment of the estimate will be small, and for large
$\gamma$, the adjustment of the estimate will be large. It is important
to choose an \emph{optimal} value for $\gamma$ in order to be able to
compare the algorithm with other algorithms; in the next section, we will
discuss this in more detail.  Finally, it remains to specify what action
to do given a particular estimate for the next return; we again define a
risk-seeking and a risk-avoiding approach, similar to eq.
\ref{eq:mamls_rs} and \ref{eq:mamls_ra} for the MA and MLS strategies:

In the risk-seeking approach, $q_{RS}(t)$ is defined as follows for
$\hat{r}_{IUR}(t)$:
\begin{equation}
  \label{eq:iur_rs}
  q_{RS}(t) =
  \begin{cases}
    q_{\mathrm{\min}} & \textrm{$\hat{r}_{IUR}(t) \le 0$} \\ q_{\mathrm{\max}} & \textrm{$\hat{r}_{IUR}(t) > 0$}
  \end{cases}
\end{equation}
and in the risk-avoiding approach, $q_{RA}(t)$ is defined as follows for
$\hat{r}_{IUR}(t)$:
\begin{equation}
  \label{eq:iur_ra}
  q_{RA}(t) =
  \begin{cases}
    q_{\mathrm{\min}} & \textrm{$\hat{r}_{IUR}(t) \le q_{\mathrm{\min}}$} \\
    \hat{r}_{IUR}(t) & q_{\mathrm{\min}} < \hat{r}_{IUR}(t) < q_{\mathrm{\max}}\\
    q_{\mathrm{\max}} & \textrm{$\hat{r}_{IUR}(t) \ge q_{\mathrm{\max}}$}
  \end{cases}
\end{equation}
where, for both definitions, $q_{\mathrm{\min}}, q_{\mathrm{\max}} \in
[0,1]$ and $q_{\mathrm{\min}} < q_{\mathrm{\max}}$.

It is important to note that reinforcement learning and the incremental
update rule are not identical; rather, reinforcement learning describes a
group of machine learning approaches and the incremental update rule is
one instance of these approaches.

We note eventually that a different representation of $\gamma$ in
eq.(\ref{eq:iur_instance}) could be used to study some aspects of the
prospect theory of decision-making. This theory takes into account that
decisions are made based on changes from a certain reference point, i.e.
humans for example decide differently for profits and for losses, as they
cognize losses twice as large as profits, \citep{kahnemanTversky79,
  KahnemanTversky92, edwards96, Takahashi-Terano03}.  This, however, is
not the target of the present investigations.

\subsubsection{Genetic Algorithm}

Genetic algorithms (GA) are a technique from the field of artificial
intelligence which finds approximate solutions to problems. Genetic
algorithms belong to the class of evolutionary algorithms. Genetic
algorithms are based on modelling solutions to a problem as a population
of chromosomes; and the chromosomes are candidate solutions to the
problem which gradually evolve to better solutions to the problem. The
following is a description of the instance of a genetic algorithm which
we apply to our scenario:

Let $j=1,...,C$ be a chromosome with population size $C$. Each chromosome
$j$ is an array of genes, $g_{jk}$ $(k=0,...,G-1)$. The values of the
genes are real numbers \citep{Michalewicz99}. In our model, each
chromosome $j$ represents a \emph{set of possible strategies} of an
agent, so the $g_{jk}$ refers to possible values for the risk propensity
$q$.

In the beginning, each $g_{jk}$ is assigned a random value:
$g_{jk}\in(q_{min},q_{max})$.  Each chromosome $j$ is then evaluated by a
\emph{fitness function}, $f_{j}(\tau)$, which is defined as follows:
\begin{equation}
\label{eq:ga_fitness}
  f_{j}(\tau)=\sum_{k=0}^{G-1} r(t)\,g_{jk} \;; \quad k \equiv t \bmod G
\end{equation}
In our model, the fitness is determined by the gain/loss that each strategy
$g_{jk}$ yields depending on the RoI, $r(t)$.  Since the fitness of a
chromosome must to be maximised, negative $r(t)$ lead to very small values of
$g_{jk}$, i.e. a low risk propensity, whereas positive $r(t)$ lead to larger
values of $g_{jk}$. This lets us consider the product of $r(t)g_{jk}$ as a
performance measure of a chromosome.

The values of $g_{jk}$ are always multiplied by different $r(t)$ values
-- i.e., depending on $t$. For the chromosome, we define a further time
scale $\tau$ in terms of generations. A generation is completed after
each $g_{jk}$ is multiplied by a RoI from consecutive time steps, $t$.
This means that the index $k$ refers to a particular time $t$ in the
following manner: $k \equiv t \bmod G$, which means $k=\hat{t} \in
\{1,G\}$, with $t=\hat{t}+ \tau G$, $\tau=0,1,2, ...$.

After time $\tau$, the population of chromosomes is replaced by a new
population of better fitting chromosomes with the same population size
$C$.  This new population is determined in the following manner: after
calculating the fitness of each chromosome according to eq.
(\ref{eq:ga_fitness}), we find the best chromosomes from the old
population by applying elitist and tournament selection of size two:
\begin{itemize}
\item \emph{Elitist selection} considers the best $s$ percentage of the
  population which is found by ranking the chromosomes according to their
  fitness. The best chromosomes are directly transferred to the new
  population.
\item \emph{Tournament selection} is done by randomly choosing two pairs
  of two chromosomes from the old population and then selecting from each
  pair the one with the higher fitness.  These two chromosomes are not
  simply transferred to the new population, but undergo a transformation
  based on the genetic operators crossover and mutation, as follows: A
  single-point crossover operator finds the cross point, or cut point, in
  the two chromosomes beyond which the genetic material from two parents
  is exchanged, to form two new chromosomes. This cut point is the
  integer part of a random number drawn from a uniform distribution
  $p_c\in U(1,G)$.
\end{itemize}
After the crossover, a mutation operator is applied to each gene of the
newly formed chromosomes. With a given mutation probability $p_{m}\in
U(0,1)$, a gene is to be mutated by replacing its value by a random
number from a uniform distribution $U(q_{min},q_{max})$. After the cycle
of selection, crossover and mutation is completed, we eventually arrive
at a new population of chromosomes that consists of a percentage of the
best fitted chromosomes from the old population plus a number of new
chromosomes that ensure further possibilities for the evolution of the
set of strategies.

Given the optimised population of chromosomes representing a set of
possible strategies, the agent still needs to update its actual risk
propensity, $q_{i}(t)$. This works as follows: at time $t=\tau$, the
agent takes the set of strategies $g_{jk}$ from the chromosome $j$ with
the highest fitness in the previous generation. Given $G=T$, this means
that the agent for each time step of the upcoming cyclic change chooses
the appropriate risk propensity by computing the following:
\begin{equation}
  \label{eq:ga} q_{GA}(t)=g_{jk} \quad \mathrm{with}\; j=\arg\,(\max\nolimits_{j=1,...,C}f_{j})\;;\;\; t \equiv k \bmod G
\end{equation}
This concludes the overview of the different agent strategies applied in
our scenario. In the following section we will adjust the respective
parameters of each of these strategies so that they are optimal. This is
crucial for a comparison of the different agent strategies -- only
strategies that perform at their best can be compared.

\section{Return on Investment}
\label{sec:ReturnOnInvestment}

Given the assumptions stated, we have to provide a function for $r(t)$
which is independent of $q(t)$. In some of the models previously studied
in the literature \citep{Kesten73}, the
influence of the market is simply treated as random, i.e. $r(t)$ is a
random number drawn from a uniform distribution in the interval $[-1,1]$.

However, it is known that for returns with a uniform distribution
centered around the origin and agents which do not have any information
on future market returns, this situation will lead the agents to a
complete loss of their budget. This is a well-known property of
multiplicative stochastic processes \citep{NavarroEtAl07}. The only way
to make a profit on the return is by having a certain knowledge on future
returns of the market.  This requires both that there exist correlations
in the market return function, and that the agents are able to resolve
and use those correlations to make correct predictions. Consequently, we
choose to introduce correlations in our market return $r(t)$ under the
form of a seasonal or periodic signal. 

\begin{figure}[htbm]
  \begin{center}
    \includegraphics[width=6.7cm]{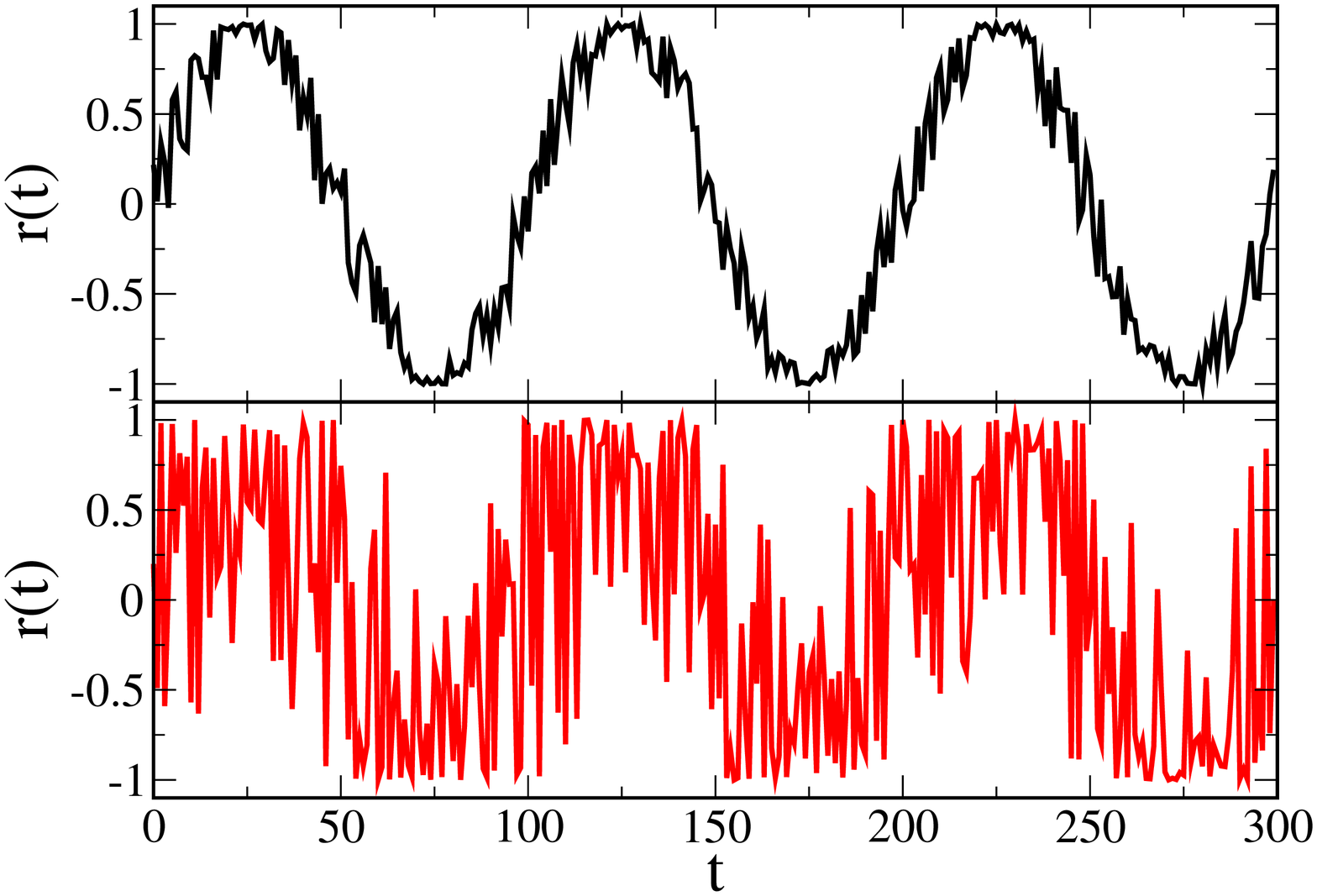}\hfill
    \includegraphics[width=6.7cm]{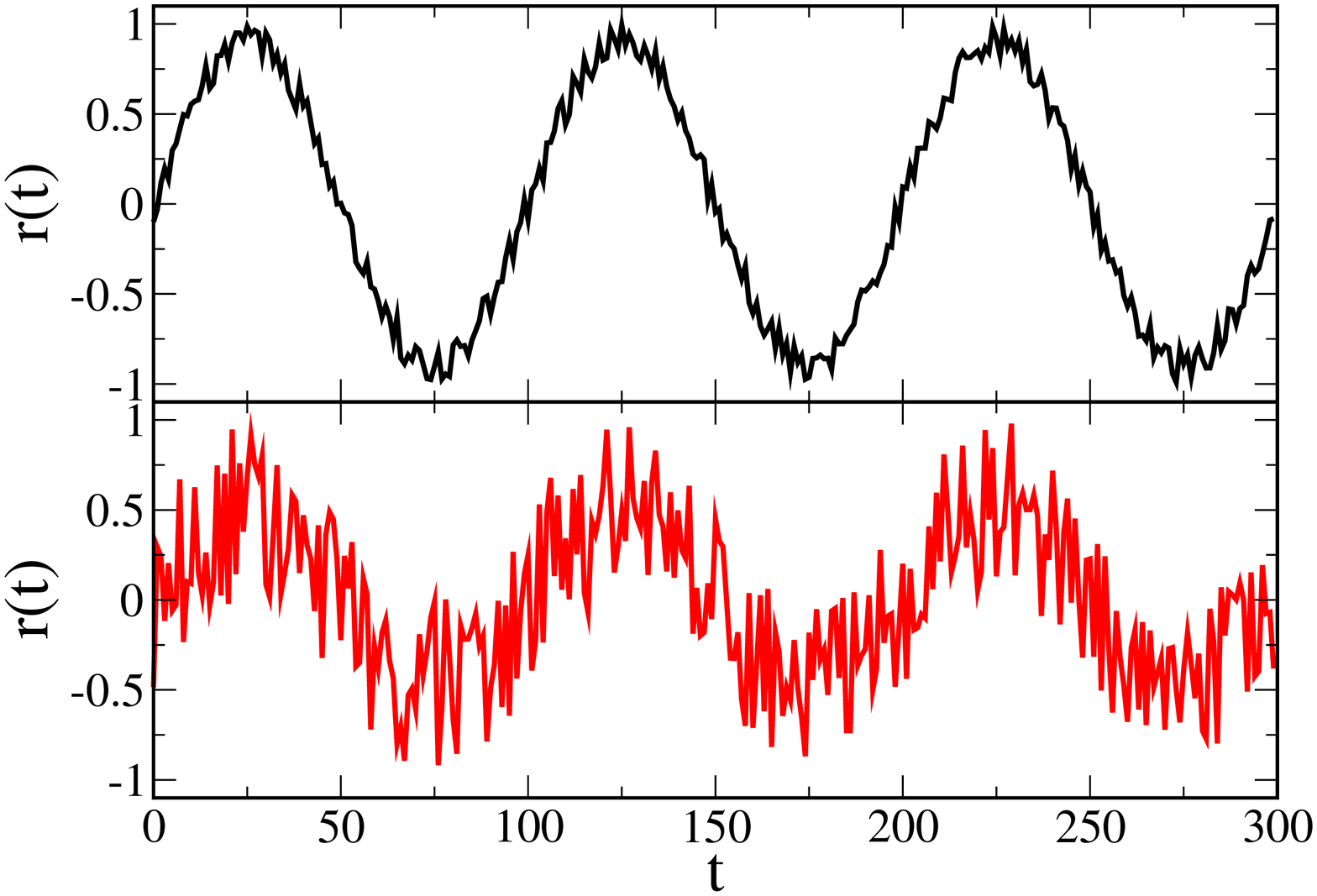}\\
    (a)\hspace{6cm} (b)
    \caption{Plots of the return functions $r(t)$, eq. \ref{eq:r} for: (a) two different phase noise levels: (top) $\sigma_1=0.1$ and (bottom) $\sigma_1=0.5$; (b) two different amplitude noise levels: (top) $\sigma_2=0.1$ and (bottom) $\sigma_2=0.5$.}  
    \label{fig:roi_sin}
  \end{center}
\end{figure}

We study two different market return functions that depend on a noise
level $\sigma_{1,2}$; for $\sigma_{1,2}=0$, they correspond to a pure
sine wave function with frequency $w$ and for $\sigma_{1,2}=1$, they are
completely uncorrelated:
\begin{eqnarray}
  \label{eq:r} 
  r_{Phase}(t) & = & \sin(w\,t+\sigma_{1}\,\pi\,\xi) \\
  r_{Amplitude}(t) & = & (1-\sigma_{2})\sin(w\,t)+\sigma_{2}\,\xi
\end{eqnarray}
where $\xi$ is distributed uniformly in the interval $[-1,1]$, i.e.  $\xi
\in U(-1,1)$. There are two types of noise that can occur with such a
sine wave function -- noise on the phase and noise on the amplitude. We
consider both cases: the first function can be seen as a periodic market
return signal with phase noise (determined by $\sigma_{1}$), the second
as a periodic market return signal with amplitude noise (determined by
$\sigma_{2}$). In our simulations, we chose the arbitrary value of
$T=100$ for the period of the sine wave.  Fig. \ref{fig:roi_sin} shows
plots for these two kinds of return functions with different noise
levels. Note that periodic returns with a periodicity changing over time
are invested recently as well.\citep{navarro-barrientos08}

The noise parameter $\sigma_{1,2}$ gives us a way of controlling the noise in
the RoI, thereby allowing us to evaluate the various strategies for different
scenarios, ranging from a completely clear signal with no noise at all (for
$\sigma_{1,2}=0$) to a noise-only signal (for $\sigma_{1,2}=1$). This makes it
possible for us to determine how well multiple strategies perform for
different types and levels of noise and what impact the type and level of
noise has on a single strategy.

Agents have no knowledge about future market returns; of course, they do not
know the functions that determine $r(t)$. Thus, the only way for agents to
maximize their gain and minimize their losses is by making a correct
prediction of the future market return and choosing the appropriate investment
action. Conceptually, we can separate an agent's strategy into two components:
a prediction component and an action component. The prediction algorithm
estimates the future values of the market return and the action algorithm
determines the best action based on the predicted results.

We study the performance of the different algorithms or strategies that are
explained in section \ref{sec:AgentStrategies}. We define the
\emph{performance} of an agent employing a particular strategy as being the
average growth of the budget $x(t)$ of the agent collected after a certain
number of time steps. We choose to take this average over $t=T=100$ time
steps, $T$ being the period of the sine wave. The reason for this particular
choice is that, for a constant investment action and a return function with no
noise, this average value will have zero standard deviation. In contrast, if
we do the averaging of the growth over all the time steps there will be a
non-zero standard deviation associated with the sine wave. In section
\ref{sec:Results}, we compare the performance of the different
strategies.

\begin{figure}[htb]
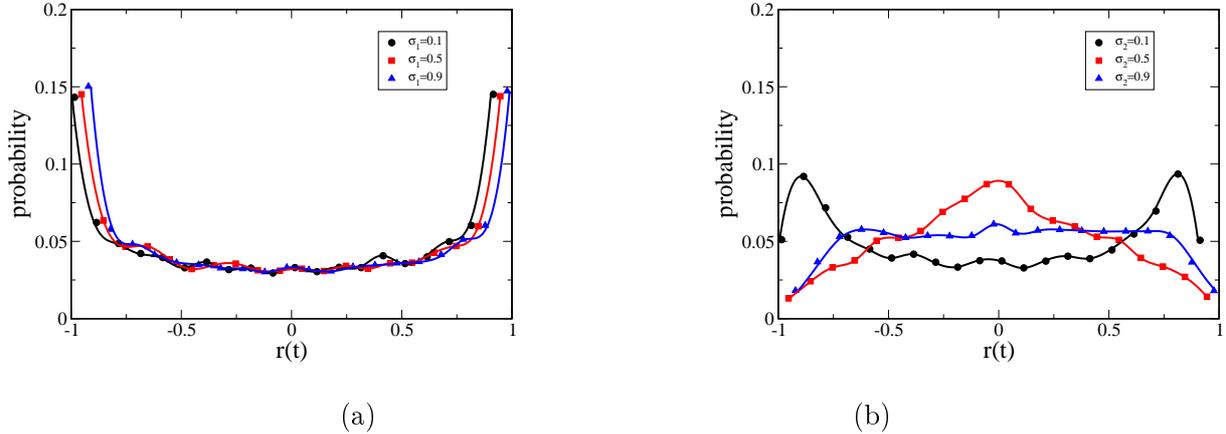

  \begin{center} 
\vspace{0.3cm}
    \includegraphics[width=6.7cm]{figures/distribution_rt_phase_noise_alternative}\hfill
    \includegraphics[width=6.7cm]{figures/distribution_rt_amp_noise_alternative}\\[0.3cm]
    (a) \hspace{6cm} (b)
    \caption{Probability distribution of the RoI, $r(t)$, eq. \ref{eq:r},
      for: (a) phase noise $\sigma_1=\{0.1,0.5,0.9\}$ and (b) amplitude
      noise $\sigma_2=\{0.1,0.5,0.9\}$.}
    \label{fig:distribution_roi}
  \end{center}
\end{figure}

In order to interpret the results that we obtain and present in this paper, it
is useful to understand some properties of the two different return functions.
Two properties are of particular interest: the absolute average value of the
return, and the correlation between the sign of two consecutive returns.

For the sake of completeness, we show the probability distribution of the
RoI in Fig. \ref{fig:distribution_roi}. For the probability distributions
of $r(t)$ with phase noise, we see that there is a higher probability for
values close to $-1$ and $1$, and a lower probability for values close to
$0$. Note that this is the same distribution that is found for a sine
wave with no noise at all. For phase noise, the value of $\sigma_1$ has
no effect -- the distributions are virtually identical. For the
probability distributions of $r(t)$ with amplitude noise, we observe a
probability distribution which is a combination of the probability
distribution of a sine wave without noise (caused by the sine wave) and a
uniform probability distribution (caused by the noise). For higher levels
of noise, the convolution of probability distributions more closely
resembles the uniform distribution, and for lower levels of noise, the
convolution of probability distributions more closely resembles the sine
wave distribution. For amplitude noise, the value of $\sigma_2$ is
crucial and different values lead to different distributions. Note that
the distribution of the returns for phase noise is independent of the
level of noise, whereas for the RoI with amplitude noise there is a
significant change as the level of noise is increased.

\begin{figure}[htb]
  \begin{center}
    \includegraphics[width=6.7cm]{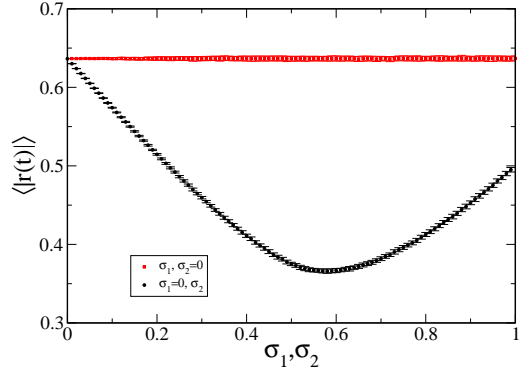}
    \caption{Average absolute value of the RoI, $r(t)$, eq. \ref{eq:r},
      for phase noise (red, top line) and for amplitude noise (black,
      bottom line).}
    \label{fig:average_abs_roi}
  \end{center}
\end{figure}

Since the distribution of the RoI is independent of the noise level $\sigma_1$
for phase noise, it is expected that the average absolute RoI, Fig.
\ref{fig:average_abs_roi}, is constant with respect to $\sigma_1$ in the RoI.
This, as we have explained, is not the case for the noise level $\sigma_2$ for
amplitude noise, where the average absolute RoI is varying with respect to
$\sigma_2$ in the RoI. For $\sigma_1=0$ and $\sigma_2=0$, the average absolute
value of the RoI are equal. Roughly, the average absolute value of the RoI
with amplitude noise decreases for $\sigma_2 < 0.6$ and it increases for
$\sigma_2 \ge 0.6$. This matches with the observations for the probability
distributions: there, for $\sigma_2=0.5$, the values are concentrated around
$r(t)=0$, leading to smaller $\mean{\abs{r(t)}}$, and for $\sigma_2=0.1$ and
$\sigma_2=0.9$, the values are less concentrated around $r(t)=0$, leading to
larger $\mean{\abs{r(t)}}$.

The average absolute RoI is of importance because its known from
multiplicative stochastic processes \citep{Sornette-Cont97,
  NavarroEtAl07}, that for a constant investment $q(t)=q_0$ the better
performing constant strategies are the ones that invest the least
possible amount. In our model, the agents are forced to invest the
minimum amount of $q_{min}=0.1$. Since $q(t)$ is multiplied with $r(t)$
in eq.  \ref{eq:budget}, the change in average absolute value of $r(t)$
has an impact similar to the change in $q_0$ seen in the multiplicative
stochastic processes \citep{NavarroEtAl07} studied. This leads to changes
in performance that are not necessarily related with the performance of
agents, and should be taken into account when interpreting the results.

\begin{figure}[htb]
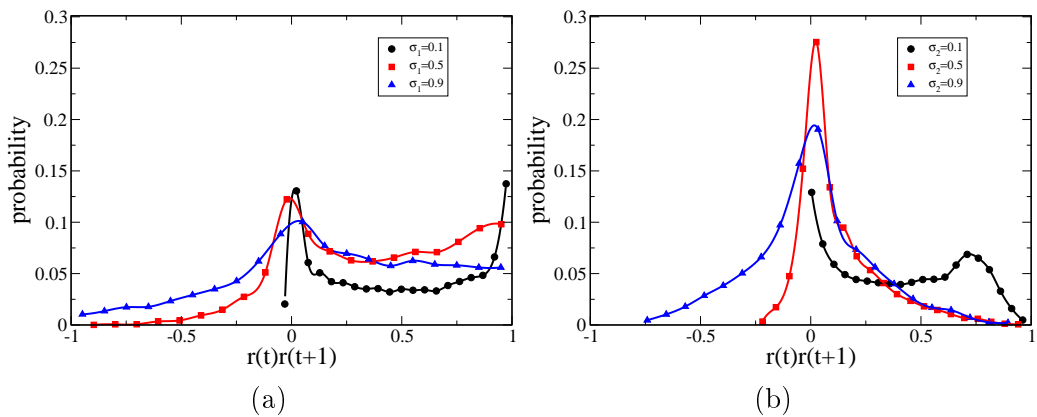

  \begin{center}
\vspace{0.3cm}
     \includegraphics[width=6.7cm]{figures/correlation_rt_phase_noise_alternative}
     \includegraphics[width=6.7cm]{figures/correlation_rt_amp_noise_alternative}
    (a)\hspace{6cm} (b)
    \caption{Distribution of the correlations of the RoI in time, $r(t)$, eq. \ref{eq:r}, for: (a) phase noise with $\sigma_1=\{0.1,0.5,0.9\}$ and (b) amplitude noise with $\sigma_2=\{0.1,0.5,0.9\}$.}
    \label{fig:correlations_roi_in_time}
  \end{center}
\end{figure}

The correlation between the sign of two consecutive returns shows whether
it is possible to draw conclusions from the sign of $r(t)$ on the sign of
$r(t+1)$. In Fig. \ref{fig:correlations_roi_in_time} we show the
distribution of the correlations of the RoI with respect to two
consecutive returns, $r(t)r(t+1)$. We can clearly see that for low levels
of noise there is bigger correlation between consecutive values. As the
noise increases this correlation diminishes until, finally, for high
levels of noise, the returns are completely uncorrelated.

Most of the algorithms studied are sensitive to correlations in consecutive
RoI with the same sign. We notice that between the returns with phase noise
and amplitude noise, correlations do not vary exactly in the same manner with
noise. In particular, it can clearly be seen that for $\sigma_{1,2}=0.5$ the
amplitude noise has still more correlation than the phase noise. This
difference can account for some discrepancies seen between the performance of
the agents for the two types of market return functions.

\section{Optimal Parameter Adjustment}
\label{sec:OptimalParameterAdjustment}

In the previous sections, we have defined several different strategies that
can be applied by agents to determine when to invest which amount of money. In
the next section, we want to compare their performance in a periodic
environment. However, in order to make this comparison meaningful, we have to
ensure that we have adjusted the different parameters of the strategies
properly. Only if the strategies perform at their optimum, they can really be
compared.

The procedure that we apply to adjust the optimal parameters is
straightforward: we compare the performance -- averaged over $10^5$ periods --
of each of the algorithms for a range of possible parameters and then choose
the optimal one. At this point, it remains to define the notion of optimality:
we have already defined that we measure the performance of agents as the
average of their budget growth over a certain number of time steps. The
optimal strategy is the strategy that performs better than all the other
strategies, i.e.  the strategy that, on average, leads to the greatest budget
growth. Of course, for the measurement, each agent has to be provided with
enough time to gather the information necessary for the proper calibration of
the algorithm that it applies.

For the MA, MLS, and IUR strategies, there is only one parameter that requires
adjustment: either the memory size $M$ (in the case of MA and MLS) or the step
size $\gamma$ (in the case of IUR). This implies that for these strategies, it
is possible to choose the optimal value of the parameter by comparing the
average budget $\mean{x(t)}$ for several possible values of the parameter, and
then take the one which gives the best results. For MA and MLS, we have
considered memory sizes $M \in [1,50]$ and for IUR, we have considered step
sizes $\gamma \in [0,1]$.
 
For the GA strategy, there are, however, several parameters which require
adjustment: the population size, $C$, the crossover probability, $p_{c}$, the
mutation probability, $p_{m}$, and the elitism size, $s$. Consequently, the
process of finding the optimal combination of values for the parameters is not
as trivial as for the other strategies. The \emph{+CARPS (Multiagent System
  for Configuring Algorithms in Real Problem Solving)} tool
\cite{Monett04a,Monett04b} was used during this step. This application uses
autonomous, distributed, cooperative agents that search for solutions to a
configuration problem, thereby fine-tuning the meta-heuristic's parameters.
The agents in \emph{+CARPS} apply a Random Restart Hill-Climbing approach and
they exchange their so-far best solutions to the problem in the process. The
intervals of definition, i.e. the intervals in which the most acceptable GA
configurations should lie, were set as follows: $C \in \{ 50, 100, 200, 500,
1000\}$, $p_{c} \in \left[0.0, 1.0\right]$, $p_{m} \in \left[0.0, 1.0\right]$,
and $s \in \left[0.0, 0.5\right]$,

\begin{table}[htb]
\label{tab:optimal_parameters}
  \caption{Optimal Strategy Parameters} 
   \centering {\begin{tabular}{|l|l|}
    \hline 
    Algorithm & Parameters \\
    \hline \hline
    MA & $M=5$ (risk-seeking), $M=2$ (risk-avoiding) \\
    \hline
    MLS & $M=25$ (both risk-seeking and risk-avoiding) \\
    \hline
    IUR & $\gamma=0.5$ (both risk-seeking and risk-avoiding) \\
    \hline
    GA & $C=1000$, $p_c=0.7$, $p_m=0.01$, $s=0.3$\\
    \hline 
  \end{tabular}}
 \end{table}
 Table 1 
 shows the optimal parameters that we choose for the comparison of the
 different strategies. Of course, the optimal parameters usually are not
 the same for different types and levels of noise or for risk-seeking and
 risk-avoiding behaviour, so at times, a compromise between several
 alternative values for different situations had to be found.

\section{Results}
\label{sec:Results}

In this section, we compare all strategies presented in this article for
RoI with periodicity $T=100$ and different noise levels for both phase
and amplitude noise.  In our comparison, we consider a set of agents,
each one using one of the following strategies: Q0 eq.~(\ref{eq:q0}), MA
eq.~(\ref{eq:ma}), MLS eq.~(\ref{eq:mls}), IUR
eq.~(\ref{eq:iur_instance}), and GA eq.~(\ref{eq:ga}). Note that periodic
returns with a periodicity changing over time are invested recently as
well.\citep{navarro-barrientos08}

In our comparison, we make two assumptions: first, all agents receive the
same RoI at a particular time, i.e. the fact that some agents win or
loose more than others is influenced only by their different strategies
to determine the correct risk-propensity value; second, all agents use
the optimal parameter values of their respective strategies. Let us state
again that only the past and current values of $r(t)$ are known to the
agents; they do not know the dynamics governing future values of $r(t)$.
 
We perform $N=100$ trials of the same experiment, i.e. RoI with same
parameters, where at each end of a cycle of the RoI, i.e. for all $t$ such
that $t{ }\bmod{ }T \equiv 0$, an average budget is obtained for each agent
over the 100 trials. This is done for a large number of time steps, i.e.
$t=10^5$. We vary the amplitude noise values, $\sigma_2\in(0,1)$, while
leaving the phase noise value constant, $\sigma_1=0$, and we vary the phase
noise values, $\sigma_1\in(0,1)$, while leaving the amplitude noise value
constant, $\sigma_2=0$. For the simulations which distinguish between a
risk-seeking and a risk-avoiding action upon a prediction of the RoI, we
compute the average budget for both approaches. This gives us four variants of
the simulations: amplitude noise/risk seeking, phase noise/risk seeking,
amplitude noise/risk avoiding, and phase noise/risk avoiding.

\subsection{Comparison}

\begin{figure}[h!]
  \begin{center} 
    \includegraphics[width=6.7cm]{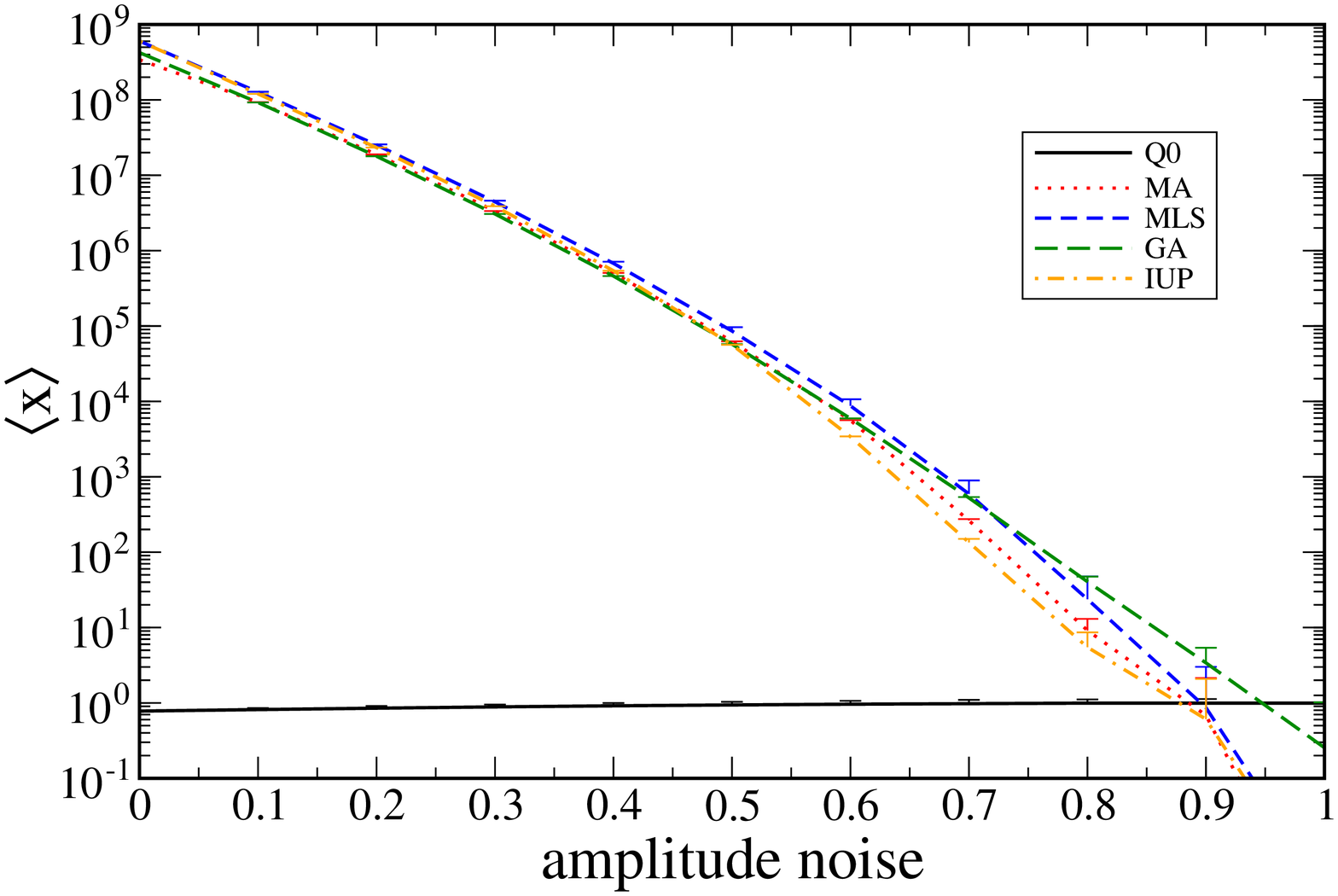}
    \includegraphics[width=6.7cm]{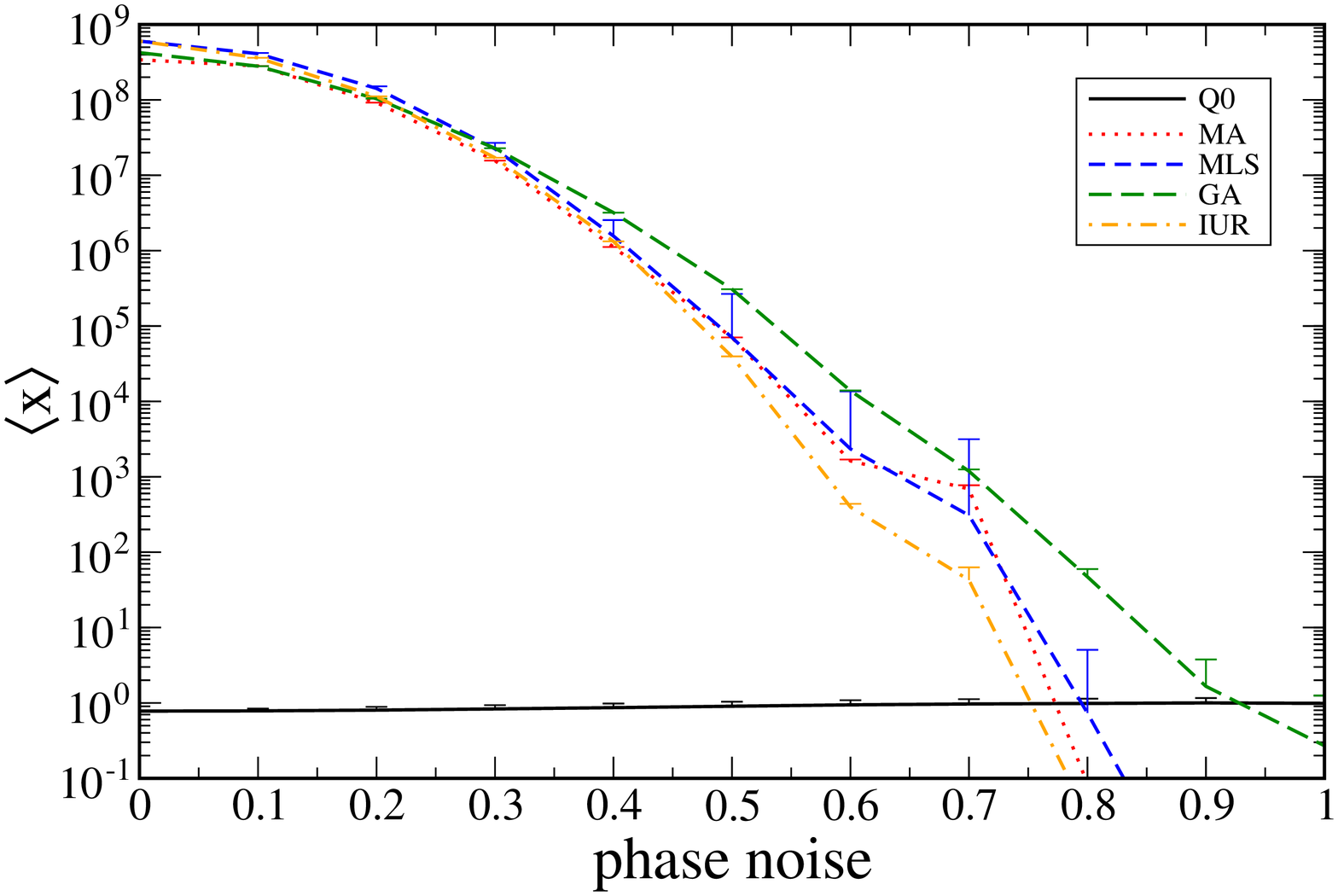} \\
    (a) \hspace{6.3cm} (b) \\
    \includegraphics[width=6.7cm]{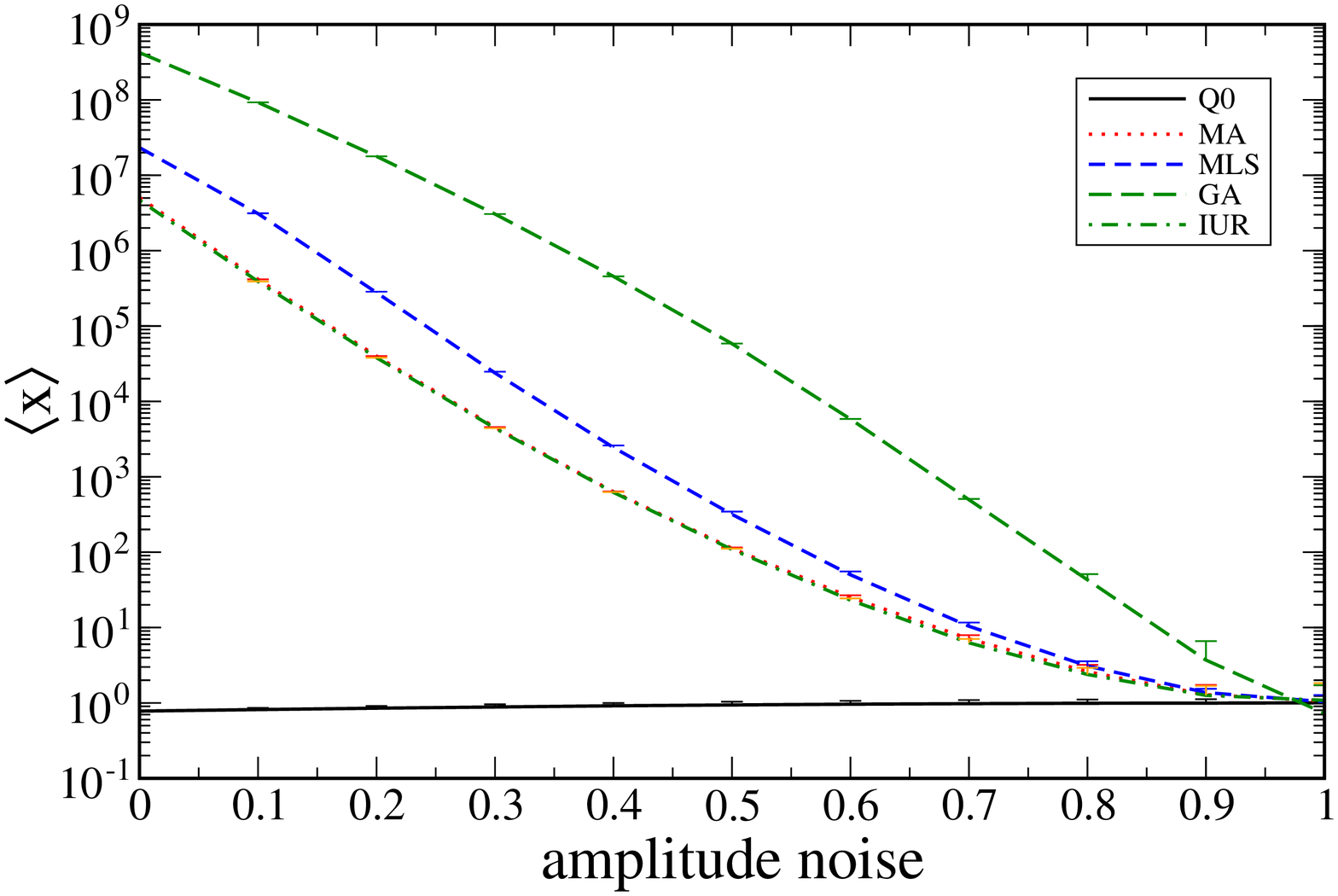}
    \includegraphics[width=6.7cm]{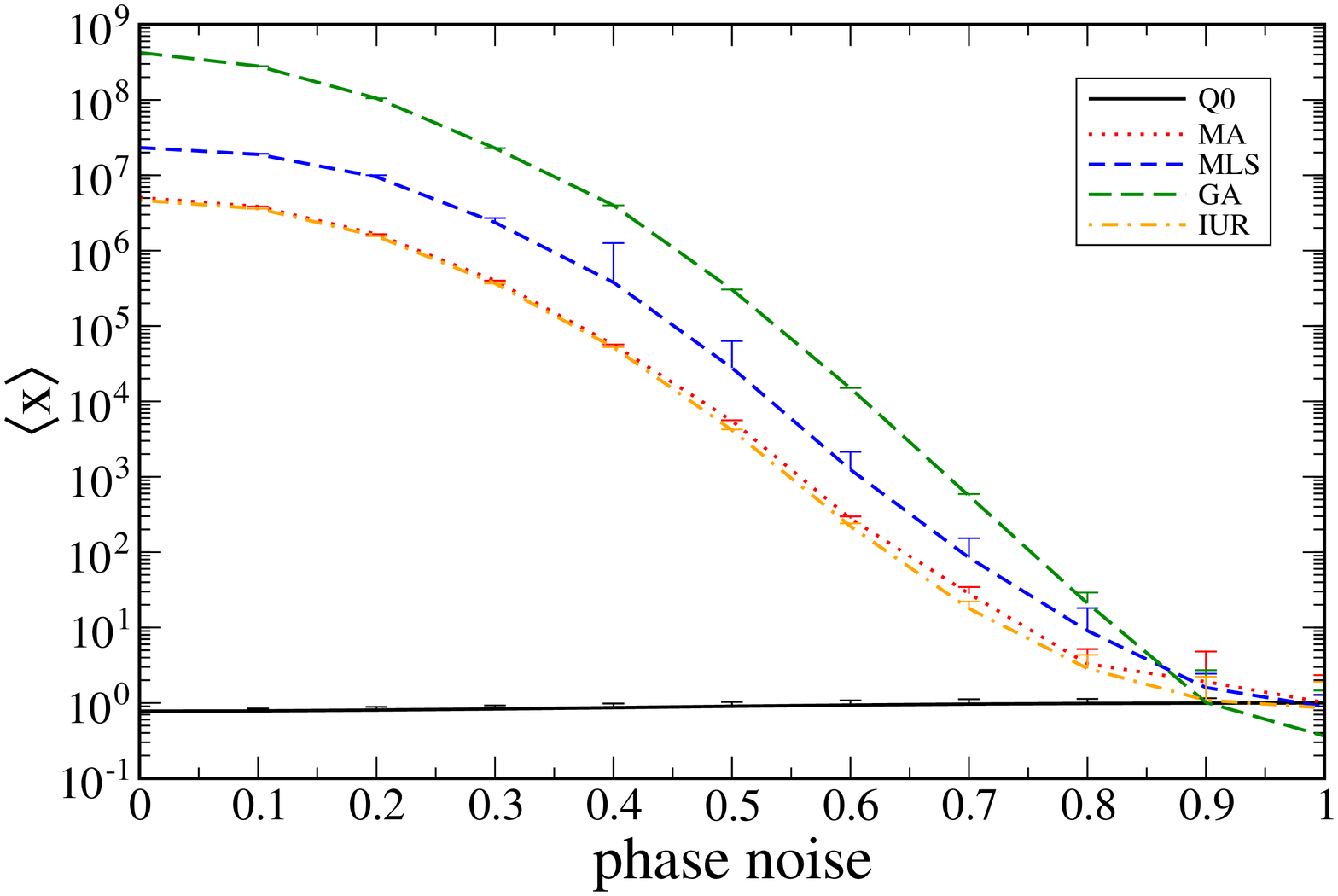} \\
    (c) \hspace{6.3cm} (d) \\
    \caption{Average budget over $N=100$ trials, for agents using
      strategies Q0, MA, MLS, GA, IUR and SW (square wave, to be
      introduced in section \ref{sec:optimalstrategy}), over $t=10^5$
      time steps. Agents use optimal parameter values for RoI with
      periodicity, $T=100$, and different noise levels: (a) different
      amplitude noise values, $\sigma_2\in(0,1)$, and no phase noise,
      $\sigma_1=0$ with a risk-seeking strategy; (b) different phase
      noise values, $\sigma_1\in(0,1)$, and no amplitude noise,
      $\sigma_2=0$ with a risk-seeking strategy; (c) different amplitude
      noise values, $\sigma_2\in(0,1)$, and no phase noise, $\sigma_1=0$
      with a risk-avoiding strategy, and (d) different phase noise
      values, $\sigma_1\in(0,1)$, and no amplitude noise, $\sigma_2=0$
      with a risk-avoiding strategy.}
    \label{fig:comparison-all}
  \end{center}
\end{figure}

Fig.~\ref{fig:comparison-all} shows the result of the simulations by plotting
the average budget resulting from the different strategies against the noise
level for each of the four variants of the simulations (amplitude noise/risk
seeking in (a), phase noise/risk seeking in (b), amplitude noise/risk avoiding
in (c), and phase noise/risk avoiding in (d)).

For all variants of the simulations, the constant-risk strategy is the
worst strategy. The constant-risk strategy always puts a constant
proportion of the budget at stake. This money is won when the return is
positive, but also lost when the return is negative; even though
$\mean{\abs{r(t)}}=0$, this leads to a loss in budget over time, as this
is a well known property for multiplicative stochastic processes.

Furthermore, for all strategies, the average budget decreases with
increasing noise. That is the expected behaviour: with increasing noise,
the accuracy of the predictions made by the agents decreases, and thus
they cannot necessarily chose the appropriate risk propensity in the
action.

There are no significant crossovers of the performance of different
strategies. In general, this implies that if a strategy $s_1$ performs
better than a strategy $s_2$ for a given noise level $\sigma_a$ (either
on the phase or on the amplitude), $s_1$ can be expected to perform
better than $s_2$ for a different noise level $\sigma_b$.  Consequently,
the \emph{choice of strategy is independent of the noise in the return}
-- a good strategy is a good strategy for all noise levels, and a bad
strategy is a bad strategy for all noise levels, too.  However, for low
noise levels, the GA is slightly outperformed by the other strategies --
this is due to the intrinsic stochastic nature of the algorithm; for the
same reason, this algorithm performs better for high noise levels.
Note that the experiments in this simulations are done for $t=10^5$ time
steps, which corresponds also to the learning phase for the GA. 

For phase noise, the average budget obtained is roughly comparable to
that for amplitude noise, although the differences between strategies are
greater for phase noise than for amplitude noise.

From the range of strategies employed, the simple strategies (MA, MLS,
IUR) were almost always outperformed by the complex one (GA). Other
researchers \cite{Gode-Sunder93, Daniels-Farmer03} have shown that this
needs not necessarily be the case.

\subsection{Optimal Strategy}
\label{sec:optimalstrategy}

\begin{figure}[htbp]
  \begin{center}
    \includegraphics[width=6.7cm]{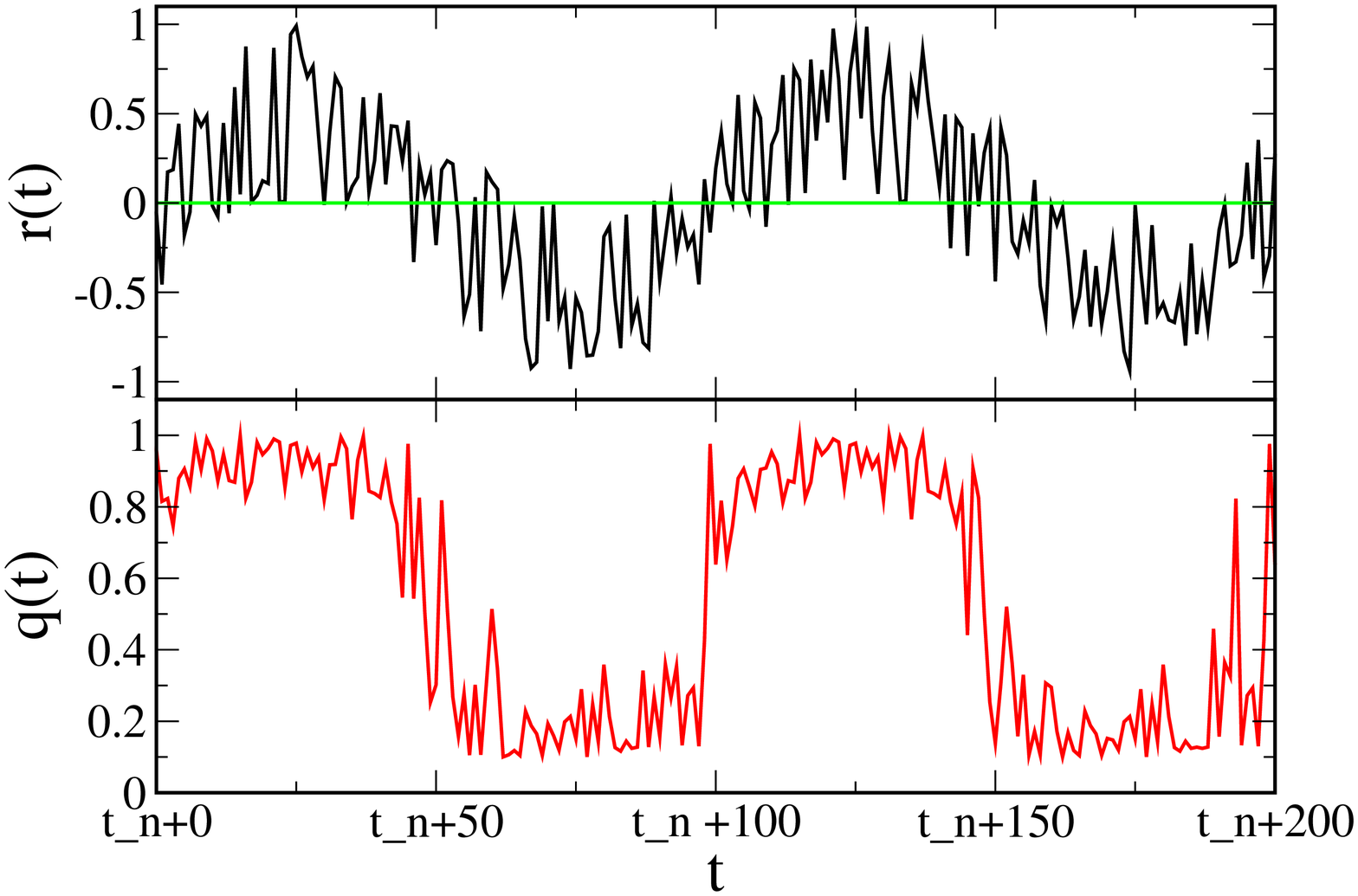}
    \includegraphics[width=6.7cm]{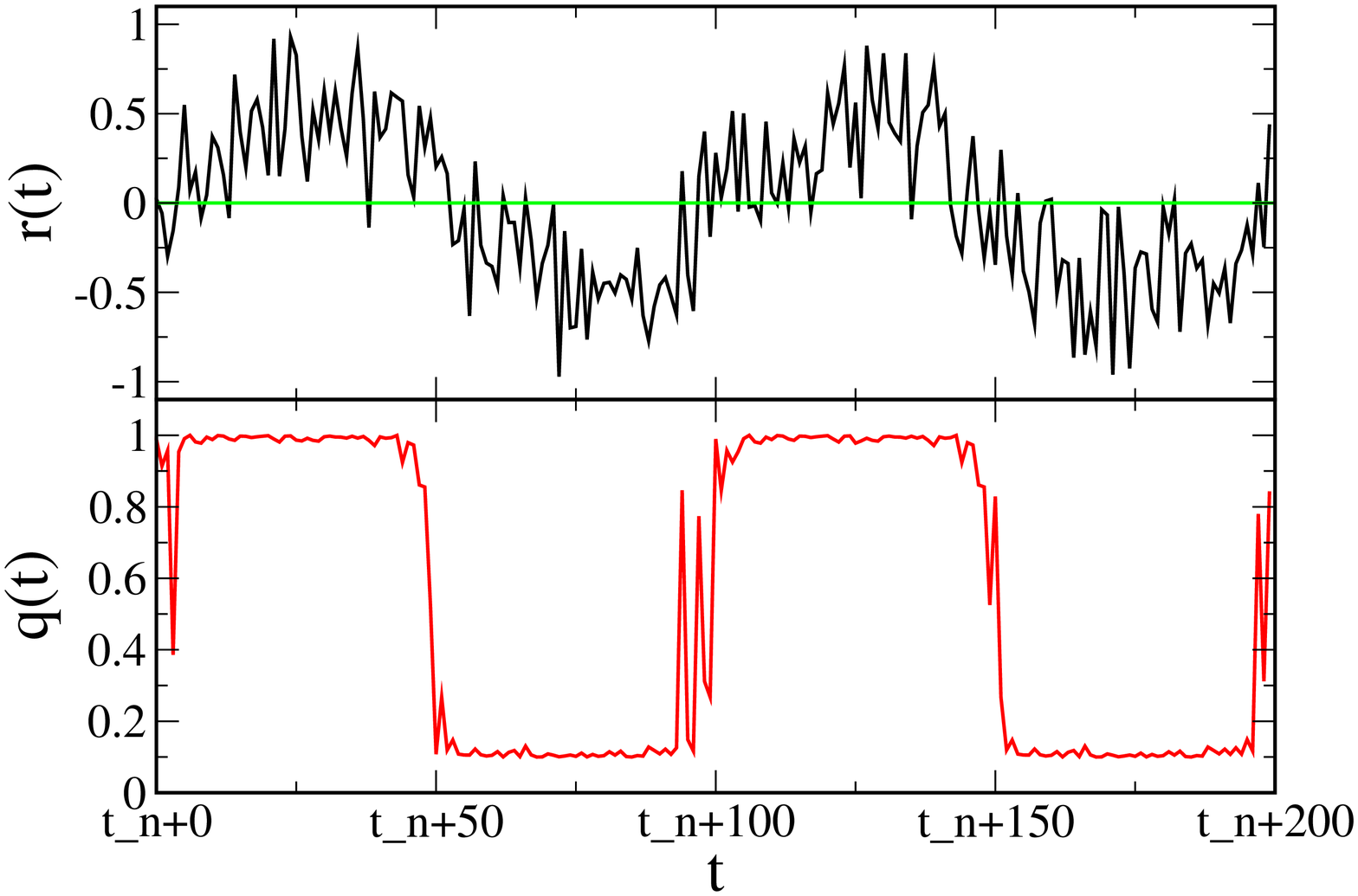} \\
    (a)\hspace{6.3cm} (b) \\
    \includegraphics[width=6.7cm]{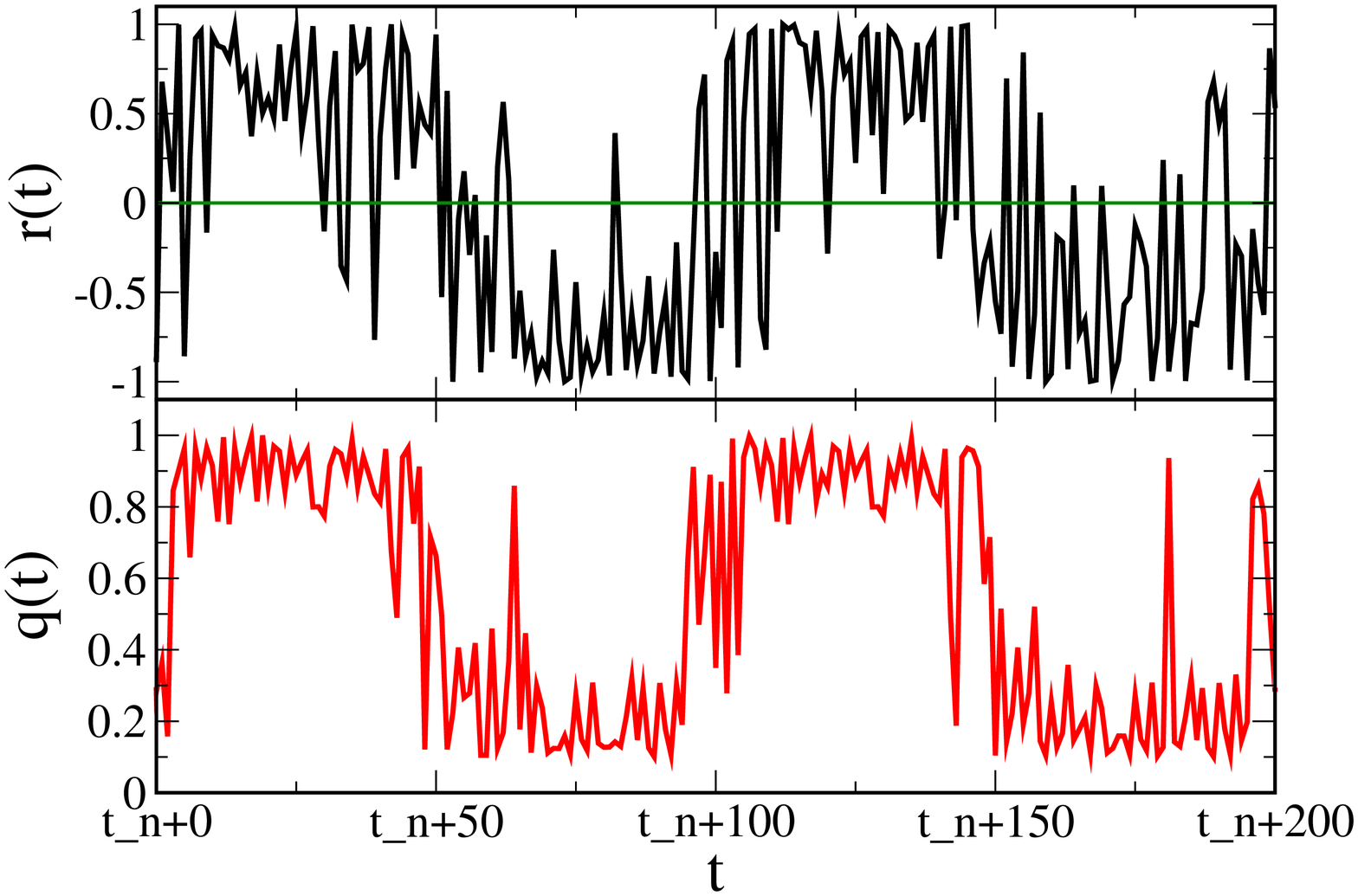}
    \includegraphics[width=6.7cm]{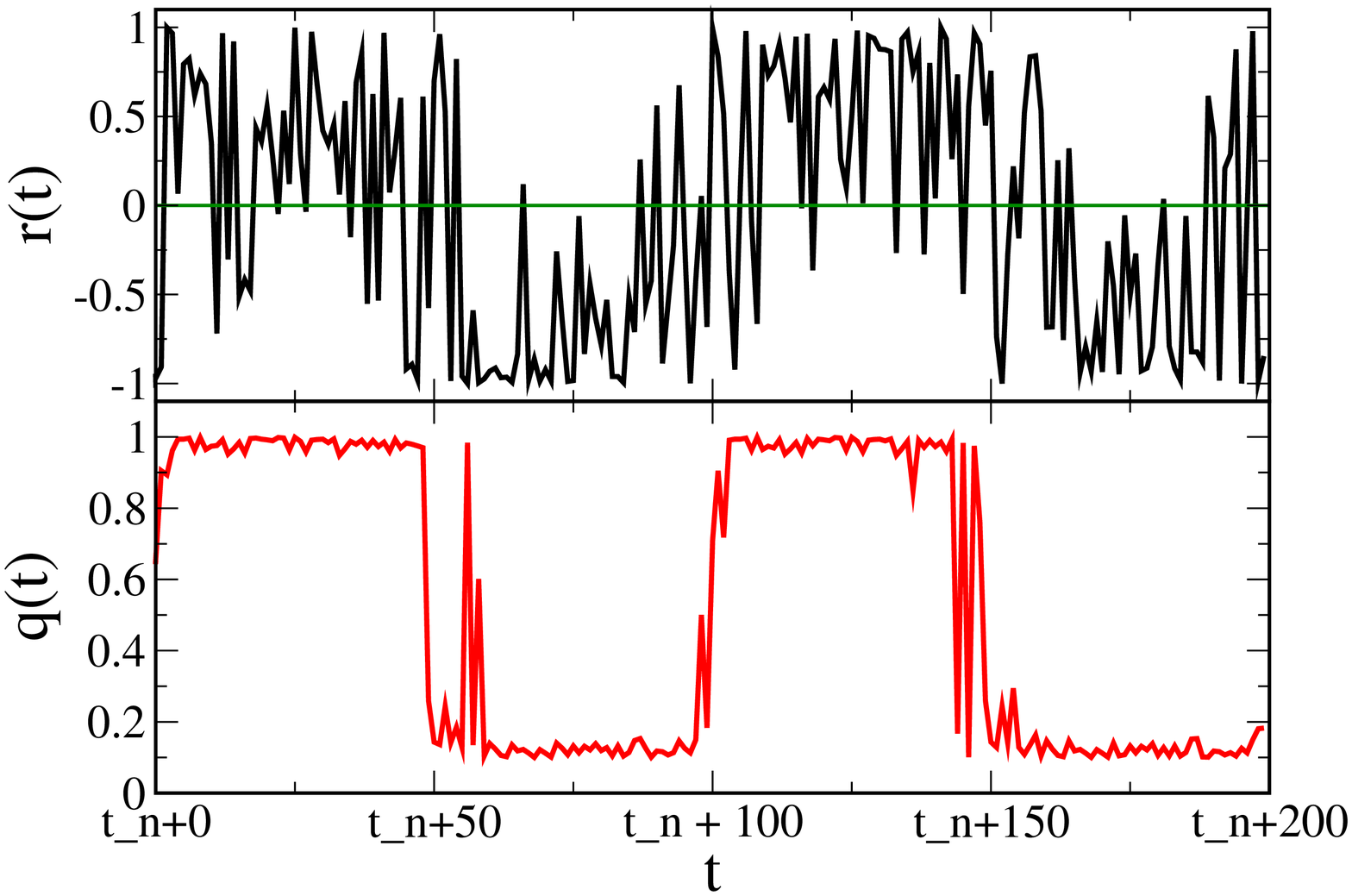} \\
    (c)\hspace{6.3cm} (d) \\
    \caption{Values of the return $r(t)$ and the risk propensity
      $q(t)$ as chosen by the GA for RoI with different types of noise
      and for different times during the simulation: (a) amplitude
      noise, $\sigma_1=0, \sigma_2=0.5$, $t_n \approx 10,000$, (b)
      amplitude noise, $\sigma_1=0, \sigma_2=0.5$, $t_n \approx
      100,000$, (c) phase noise, $\sigma_1=0.5, \sigma_2=0$, $t_n
      \approx 10,000$, (d) phase noise, $\sigma_1=0.5, \sigma_2=0$,
      $t_n \approx 100,000$.}
    \label{fig:ga_sw_behaviour}
  \end{center}
\end{figure}

As a consequence of the comparison it is logical to investigate what would be
the optimal strategy in the given scenario. Given the fact that the GA
performs best of all the strategies, it makes sense to look at the $q(t)$ as
chosen by the GA for $r(t)$ over time, in order to analyse why the GA performs
so well. Fig. \ref{fig:ga_sw_behaviour} plots the values of $r(t)$ and the
corresponding $q(t)$ as chosen by the GA against time $t$ for different noises
and from different times $t_n$ on. From the graph, it is visible that the
behaviour of the GA resembles a square wave function which is a type of a
ramp-rectangle function.

The ramp-rectangle (RR) function maps RoI that are uncertain to
increasing/decreasing risk-propensity values and RoI that are certainly
positive or negative to a maximal or minimal risk-propensity value,
respectively. The corresponding strategy is expressed as follows:
\begin{equation}
  \label{eq:rr}
  q(t+1) =
  \begin{cases}
    \left(\frac{q_{\mathrm{max}}-q_{\mathrm{min}}}{h_1}\right)\hat{t}+q_{\mathrm{min}} &  \textrm{if $\hat{t}\in(0,h_1)$} \\
    q_{\mathrm{max}} & \textrm{if $\hat{t}\in[h_1,h_2]$} \\
    \left(\frac{q_{\mathrm{max}}-q_{\mathrm{min}}}{h_2-h_3}\right)(\hat{t}-h_2)+q_{\mathrm{max}} &  \textrm{if $\hat{t}\in(h_2,h_3)$} \\
    q_{\mathrm{min}} & \textrm{if $\hat{t}\in[h_3,h_4]$}
  \end{cases}
\end{equation}
In this function, $h_1$ ($h_3$) sets the transition from an increasing
(decreasing) ramp function to a rectangle function and $h_2$($h_4$) sets the
transition from a rectangle function to a decreasing (increasing) ramp
function. Moreover, for each time step, $t$, the following congruence is used:
$\hat{t} \equiv t \bmod h_4$; this maps each time step $t\in(0,\infty)$ to a
time step in the ramp-rectangle function, $\hat{t}\in(0,h_4)$.

Furthermore, we assume that the differences between time steps when an agent
increases and decreases its risk propensity values are symmetric. This means
that the time difference $\triangle h$ between when the ramp function starts
and stops to increase or decrease can be expressed as follows:
\begin{equation}
  \label{eq:rr_difference}
  \triangle h=h_1=h_3-h_2
\end{equation}
which for $\triangle h=1$, means that agents use a \emph{Square Wave} (SW)
strategy. We are particularly interested in this case of the ramp-rectangle
function: it implies that an agent invests $q_{max}$ for time steps
$\hat{t}\in(0,T/2)$, and invests $q_{min}$ for time steps $\hat{t}\in[T/2,T]$.
This is the optimal strategy.

The GA approaches the optimal strategy: for all different noises, the
risk propensity $q(t)$ chosen by the GA approximates the one that would
have been chosen by SW. Considering that the GA does not have an `a
priori'-behaviour defined, it is interesting to realise that it finds the
\emph{optimal} strategy -- investing the maximum when, at a particular
time $t$ in the period, the probability of winning is higher than loosing
and vice versa -- on its own.

Fig. \ref{fig:ga_sw_behaviour} illustrates this behaviour. It plots the
values of $r(t)$ and the corresponding $q(t)$ as chosen by the genetic
algorithm against time $t$ for different noises and from different times
$t_n$ on. From this, it is clearly visible that the behaviour of the GA
is very similar to the behaviour of the SW, which is the optimal
strategy.  Comparing fig.~\ref{fig:ga_sw_behaviour} (a) with (b) and
fig.~\ref{fig:ga_sw_behaviour} (c) with (d), i.e. the same scenario, but
at different times $t_{n1}=10,000$ and $t_{n2}=100,000$, one can see that
the $q(t)$ chosen by the GA approach the ones chosen by the SW more
closely as time goes on -- i.e., as the GA has more time to evolve.
Additionally, from the simulation results, it can be observed the
approximation of the SW by the GA is closer for low levels of noise than
for high levels of noise. This is the expected behaviour.

\subsection{``Everything on Red''}

Furthermore, from the plots, we can observe that for low levels of noise,
the \emph{risk-seeking} behaviour clearly outperforms the
\emph{risk-avoiding} behaviour: always investing the maximum when a
positive return is expected and investing the minimum when a negative
return is expected outperforms investing a quantity proportional to the
expected return. This may seem counter-intuitive -- humans would probably
choose not to invest their complete budget when they know that there is a
certain probability to lose it.

\begin{figure}[htbp]
  \begin{center}
    \includegraphics[width=6.7cm]{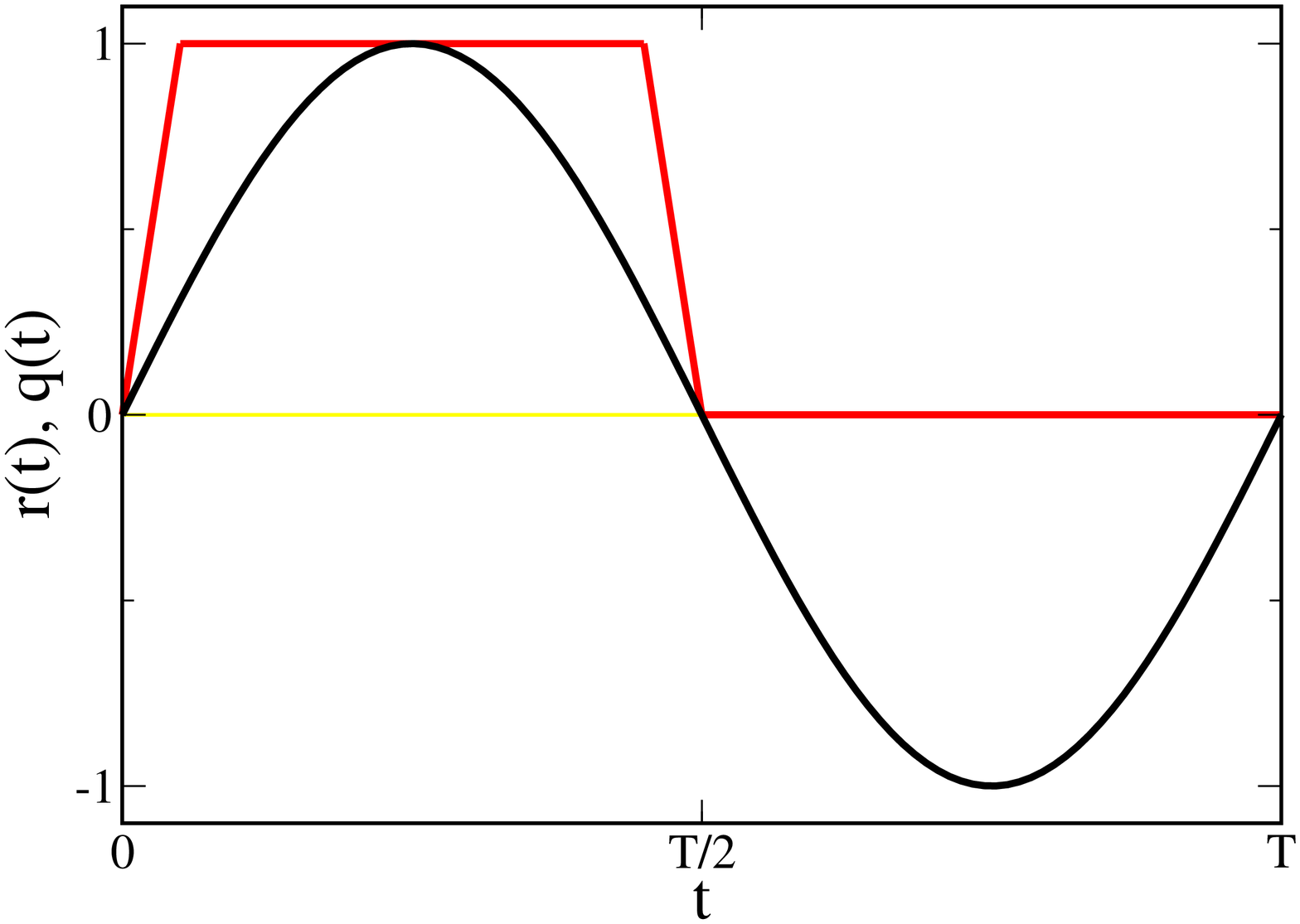}
    \includegraphics[width=6.7cm]{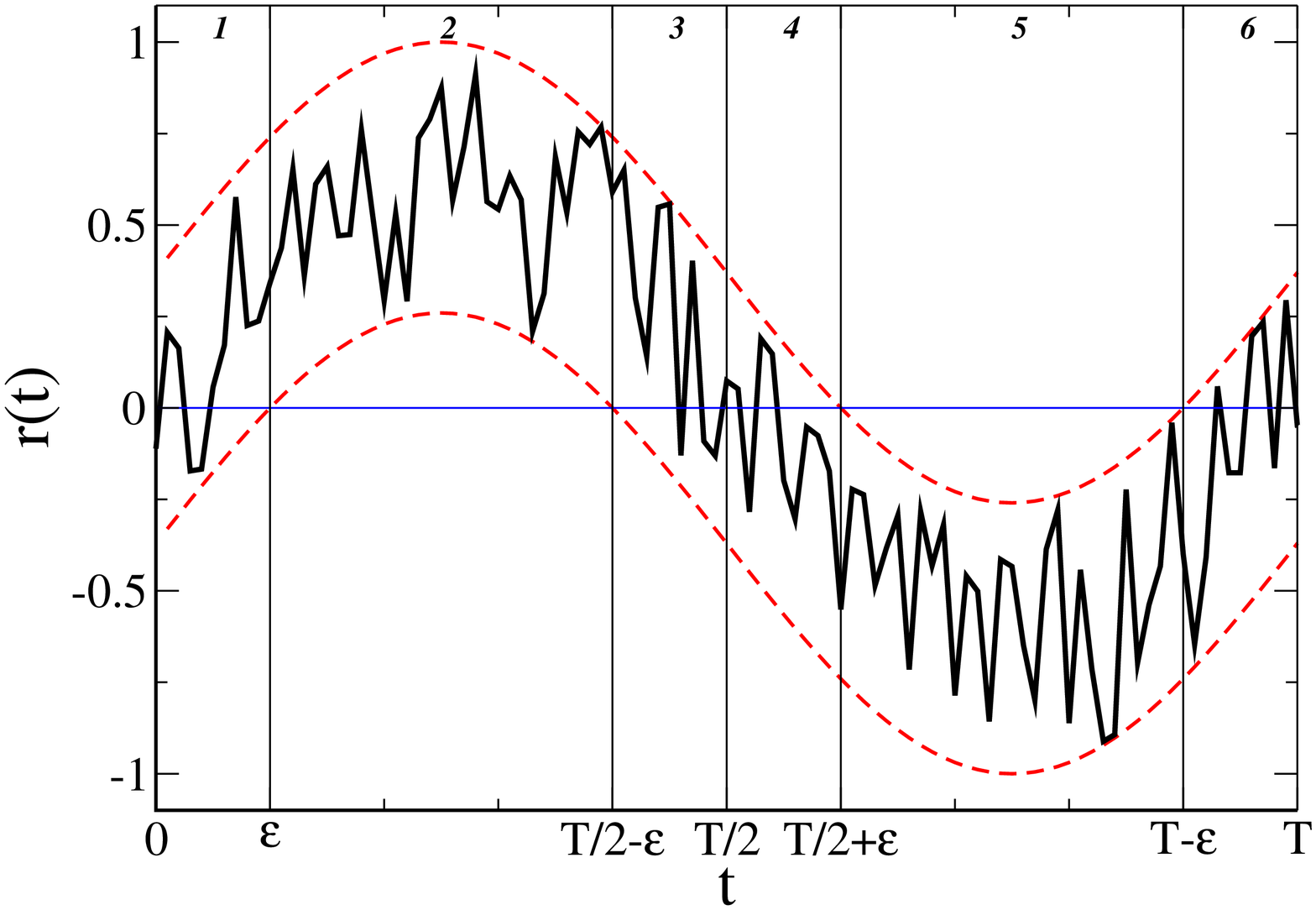}
    (a)\hspace{6cm} (b)
    \caption{Intervals of certainty and uncertainty: (a) shows $r(t)$
      with no noise and the corresponding $q(t)$ of the square wave (SW)
      strategy plotted against $t$ and (b) shows $r(t)$ with noise and
      the different intervals for which different conclusions about the
      sign of the return can be drawn: (2) and (5) are the intervals in
      which the sign of $r(t)$ is certain to be positive or negative,
      respectively, and (1), (3), (4), and (6) are the intervals in which
      the the sign of $r(t)$ is uncertain.}
    \label{fig:certain_uncertain_intervals}
  \end{center}
\end{figure}

However, this behaviour is explained as follows: consider $r(t)$ to be
periodic with a period of $T$ and, for the moment, assume that there is
no noise, i.e. $\sigma_1=0$ as well as $\sigma_2=0$. Then, the optimal
strategy would be to invest the complete budget or $q_{max}$ during
$[0,T/2)$ and to invest nothing or $q_{min}$ during $[T/2,T)$.  This is
because it is certain -- we assumed that there is no noise -- that during
the first half of the period, $[0,T/2)$, the value of $r(t)$ will be
positive and during the second half of the period, $[T/2,T)$, the value
of $r(t)$ will be negative. No matter what the precise values of $r(t)$
are, once they are positive, this leads to a gain, and thus $q(t)$ should
be as large as possible to maximise the gain; conversely, once the values
of $r(t)$ are negative, this leads to a loss, and thus $q(t)$ should be
as small as possible or zero to minimise the loss. In other words,
\emph{for determining $q(t)$, not the quantity of the expected return
  matters, but whether the probability of the expected return being
  positive is greater than the probability of the expected return being
  negative}. This explains why the risk-seeking behaviour outperforms the
risk-avoiding behaviour for periodic returns with no noise. The behaviour
of this strategy is shown in Fig.~\ref{fig:certain_uncertain_intervals} (a).

For periodic returns with noise, i.e., $\sigma_1 \ne 0$ or $\sigma_2 \ne
0$, the situation is quite similar. Depending on the values of $\sigma_1$
and $\sigma_2$, there will be two intervals $[0+\epsilon,(T/2)-\epsilon)$
and $[(T/2)+\epsilon,T-\epsilon)$ such that during
$[0+\epsilon,(T/2)-\epsilon)$, the value of $r(t)$ will -- on average --
be positive and such that during $[(T/2)+\epsilon,T-\epsilon)$, the value
of $r(t)$ will -- on average -- be negative, see
Fig.~\ref{fig:certain_uncertain_intervals} (a). In these intervals, the
optimal strategy would again be to invest the complete budget or
$q_{max}$ and to invest nothing or $q_{min}$, respectively.  The value of
$\epsilon$, of course, depends on $\sigma_1$ and $\sigma_2$, i.e. the
more noise, the greater $\epsilon$. Now, what still has to be considered
are the intervals $[0,0+\epsilon)$, $[(T/2)-\epsilon,(T/2)+\epsilon)$,
and $[T-\epsilon,T)$. Because of the noise, it is not possible to
determine the exact sign of $r(t)$ during these intervals.

However, it still is possible to say that -- on average -- the probability of
$r(t)$ being positive is greater than the probability of $r(t)$ being negative
during $[0,0+\epsilon)$ and $[(T/2)-\epsilon,T/2)$ and the probability of
$r(t)$ being negative is greater than the probability of $r(t)$ being positive
during $[T/2,(T/2)+\epsilon)$ and $[T-\epsilon,T)$. Consequently, it makes
sense to invest during $[0,0+\epsilon)$ and $[(T/2)-\epsilon,T/2)$ and not to
invest during $[T/2,(T/2)+\epsilon)$ and $[T-\epsilon,T)$.

With such behaviour, there will, however, be the situation that an agent
invests the complete budget, but the return is negative. In this type of
situation, $\abs{r(t)}$ depends on $\sigma_1$ and $\sigma_2$: for small
$\sigma_1$ and $\sigma_2$, it will be small, too. Consequently, for low
levels of noise, the product of $q(t)r(t)$ would be a small value, which
signifies, for $r(t)<0$, a small loss, and for $r(t)>0$, a small gain.
Thus, even for $q(t)=1$, the loss is bound to a proportion of the budget
corresponding to the value of $r(t)$.  This explains why the risk-seeking
behaviour outperforms the risk-avoiding behaviour for low levels of
noise. For high levels of noise, the product of $q(t)r(t)$ needs not
necessarily be a small value, which potentially signifies, for $r(t)<0$,
a large loss, and for $r(t)>0$, a large gain. Thus, an agent could
potentially loose a significant amount of its budget if it invests the
complete budget; this is the reason why, for \emph{high} levels of noise,
the risk-avoiding behaviour \emph{outperforms} the risk-seeking
behaviour.

This also provides a straightforward explanation why different algorithms
using the same rule to determine $q(t)$ perform differently. Even though the
best strategy is to still invest the maximum when there is a slightly better
probability that $r(t)>0$ than that $r(t) \le 0$, the algorithms fail to
predict the exact probabilities of $r(t)>0$ and of $r(t) \le 0$ with good
enough accuracy to determine how to properly invest. In other words, the
performance of the action depends on the accuracy of the prediction; if the
accuracy of the prediction is high, then the performance of the action is
good, and if the accuracy of the prediction is low, then the performance of
the action is bad, too. The GA does not exhibit this prediction-action
behaviour and it is able to adjust better than the other strategies.

\section{Discussion and Conclusions}
\label{sec:Conclusions}

In this paper, we have presented a number of strategies that can be
applied by agents in an investment market scenario with periodic returns
and different types and levels of noise. We have compared their
performance -- the respective average budget growth over a certain number
of time steps -- and analysed the results.  We have made three main
observations:
\begin{enumerate}
\item The \emph{type} of noise -- whether the RoI has phase or amplitude
  noise -- does not have a significant influence on the performance of
  the algorithms, while the \emph{level} of noise certainly does -- for
  increasing noise, we observe decreasing performance.
\item The GA performs best of all strategies for almost all scenarios; it
  discovers a strategy which resembles a square wave strategy and which
  follows the principle of always investing the complete budget or the
  maximum amount possible when the expected return is positive and not
  investing anything or the least amount possible when the expected
  return is negative.
\item The best rule for investment is the \emph{risk-seeking behaviour}
  of always putting the complete budget in an investment; this behaviour
  clearly outperforms a risk-avoiding behaviour which humans would
  probably apply intuitively: whilst it may seem intuitive to a human to
  invest an amount proportional to the expected return, this is not the
  approach which yields the greatest budget growth over time.
\end{enumerate}

Consequently, returning to our original goal to find an answer to the
question of to which extent the internal complexity of agents influences
their overall performance, we can state that, in our simple scenario, the
agents with a complex architecture outperform the agents with a simple
architecture.

Although, with respect to the question above, the major focus on this
paper is more related to issues of computer science, one may also ask
for the application of the results in an economic context, in particular
to financial markets. Surely, our paper can be seen as a computational
experiment on the performance of different \emph{trading} strategies in
face of noisy market returns. In this context, the agent in our model may
have two possible preferences: liquidity preference \citep{tobin58} and
speculative preference. I.e. based on the previous returns the agent has
a preference to keep cash or to invest in the market, respectively -
which is modeled by the risk-seeking and risk-avoiding behavior.

Apart from this, our model allows only a limited interpretation in the
context of financial markets, because a number of important features in
these markets are not covered or are even explicitely excluded, for the
sake of a controlled simulation setup:

\begin{description}
\item[\emph{Heterogeneity}:] Agents in our model are homogenous with
  respect to the strategy employed, i.e. there is no variability in their
  individual strategies.\citep{kirman05} In this respect, the model is
  basically a ``representative agent model'', which takes into account
  only the limited information of the previous $r(t)$. More elaborated
  strategies, where agents are assumed to be fundamentalist or chartists
  \citep{day90,farmer00price,follmer05} are also not considered here.

\item[\emph{Interaction}:] Agents in our model do not interact with other
  agents. They rather ``learn'' the dynamics of the market return, in
  order to predict it more accurately. Important collective interactions
  in financial markets, such as herding behavior, is neglected here, as
  well as interactions between (heterogeneous) trading strategies
  \citep{follmer05}.  This implies the absence of emergent properties in
  our model, as \textit{heterogeneity} and \textit{interaction} are
  indeed basic premises for the existence of emergent properties in
  financial markets.

\item[\emph{Feedback}:] Agents in our model have no effect upon the
  market, and consequently the price of an asset and the return on
  investment are treated as exogenous variables.  This is equivalent to
  the 'atomistic market' assumption.  Our model also neglects the
  collective impact of \emph{all} agents on the price and the return of
  an asset. Other feedbacks on the market, such as agent's expectations
  about the market dynamics itself, are also not explicitely modeled
  here. Some artificial market models consider an endogeneous approach,
  where the returns are generated by means of constant trading between
  heterogeneous agents \citep{day90,farmer00price}.

\item[\emph{Microfoundation}:] Our model is lacking an adequate economic
  microfoundation of the (representative) agent behaviour. The terms
  ``risk-avoiding'' and ``risk-seeking'' are used to denote the
  investment preference of the agent. However, since decisions are always
  taken based on just the expected return, the behaviour of the agent has
  to be classified as ``risk neutral'' -- risk-adverse agents indeed
  account also for the ``variance'' of returns in their decisions.
  Recent literature in economics and finance presents a more realistic
  approach about behaviour toward
  risk. \citep{hens06,Pindyck-Rubinfeld} 

\item[\emph{Multi-assets}:] Agents in our model can only invest in one
  (risky) asset, whereas in financial markets multi-asset investments and
  portfolio strategies play the most crucial role.
  \citep{elton03,markowitz91} Multi-asset optimal investment strategies
  for risky assets were already discussed 50 years ago, with an
  interesting relation to gambling \citep{breiman60}.  More recently,
  investment strategies to readjust portfolios \citep{merton90} have been
  extended \citep{maslov98} for a general distribution of return per
  capital.  Similar to our model, these contributions consider exogeneous
  returns which are drawn from a probability distribution or are modeled
  by a stochastic processes.
\end{description}

\subsection*{Acknowledgements}
\label{sec:Acknowledgements}

We thank Dagmar Monett D\'iaz for her help in finding the optimal
parameters for the genetic algorithm applied in this paper using the
\emph{+CARPS} tool and shown in Table 1. We furthermore thank Jo\~ao F.
M. Rodrigues, Mauro Napoletano and Robert Mach for valuable comments and
discussions, and the latter one for his help in obtaining the data and
creating the plots for Figure \ref{fig:rq}.

\end{document}